\documentclass[12pt]{article}

\pdfoutput=1
\usepackage{color}
\usepackage{epsfig, palatino}
\usepackage{pstricks,pst-node,pst-tree}
\usepackage{epic}
\usepackage{mathrsfs}
\usepackage{ae} 
\usepackage[T1]{fontenc}
\usepackage[ansinew]{inputenc}
\usepackage{amsmath}
\usepackage{amssymb}
\usepackage{graphicx}
\usepackage{ulem}
\usepackage{color}
\definecolor{darkblue}{cmyk}{0.9,0.9,0,0}
\usepackage[colorlinks=true,linkcolor=darkblue,citecolor=darkblue,urlcolor=darkblue]{hyperref}
\usepackage{cite}
\usepackage{hyperref}
\usepackage{wasysym}
\usepackage{varioref}
\usepackage{makeidx}
\usepackage[english]{babel}
\usepackage{simplewick}
\usepackage{array}
\usepackage{multirow}

\usepackage{animate}
\usepackage[font={small}]{caption}

\newcommand{\comment}[1]{}

\newcommand{\beq}{\begin{equation}}
\newcommand{\eeq}{\end{equation}}
\newcommand{\beqq}{\begin{equation*}}
\newcommand{\eeqq}{\end{equation*}}
\newcommand\beqa{\begin{eqnarray}}
\newcommand\eeqa{\end{eqnarray}}
\newcommand\beqaa{\begin{eqnarray*}}
\newcommand\eeqaa{\end{eqnarray*}}
\newcommand\bea{\begin{array}}
\newcommand\eea{\end{array}}

\newcommand{\nn}{\nonumber}

\newcommand{\neqa}{\nonumber\end{eqnarray}} 
\newcommand{\la}[1]{\label{#1}}

\renewcommand{\d}{\partial}

\newcommand{\<}{{\langle}}
\renewcommand{\>}{{\rangle}}

\newcommand{\re}{\relax{\rm I\kern-.18em R}}

\renewcommand{\sp}{p\hspace{-.40em}/}

\definecolor{darkgreen}{rgb}{0.0, 0.45, 0.0}

\def\XXint#1#2#3{{\setbox0=\hbox{$#1{#2#3}{\int}$}
\vcenter{\hbox{$#2#3$}}\kern-.5\wd0}}

\def\su2{{SU(2)}}

\def\a{{\alpha}}

\def\[{\left[}
\def\]{\right]}

\def\s{\sigma}
\def\a{\alpha}

\def\th{\theta}

\def\({\left(}
\def\){\right)}
\def\[{\left[}
\def\]{\right]}

\def\<{\langle}
\def\>{\rangle}

\def\i2{\frac{i}{2}}

\def\spi{\relax{\rm \pi\kern-0.5em /}}
\def\sA{\relax{\rm A\kern-0.5em /}}
\def\sp{\relax{\rm p\kern-0.5em /}}
\def\sd{\relax{\rm \d\kern-0.5em /}}
\def\sk{\relax{\rm k\kern-0.5em /}}
\def\sn{\relax{\rm n\kern-0.5em /}}
\def\sl{\relax{\rm l\kern-0.5em /}}
\def\sP{\relax{\rm P\kern-0.7em /}}
\def\sBethe{\relax{\rm \Bethe\kern-0.5em /}}

\def\2F1{\,_2{\rm F}_1}

\makeindex

\newcommand\blfootnote[1]{%
  \begingroup
  \renewcommand\thefootnote{}\footnote{\hspace{-6mm}#1}%
  \addtocounter{footnote}{-1}%
  \endgroup
}

\begin{document}

\thispagestyle{empty}

\renewcommand{\thefootnote}{\fnsymbol{footnote}}
\setcounter{page}{1}
\setcounter{footnote}{0}
\setcounter{figure}{0}

\begin{flushright}
CERN-TH-2016-163
\end{flushright}
\vspace{-0.4in}
\begin{center}
$$$$
{\Large\textbf{\mathversion{bold}
The S-matrix Bootstrap II: \\Two Dimensional Amplitudes
}\par}
\vspace{1.0cm}

\textrm{Miguel F. Paulos$^\text{\tiny 1}$, Joao Penedones$^\text{\tiny 2,\tiny 3}$, Jonathan Toledo$^\text{\tiny 4}$, Balt C. van Rees$^\text{\tiny 5}$, Pedro Vieira$^\text{\tiny 4,\tiny 6}$}
\blfootnote{\tt  \#@gmail.com\&/@\{miguel.paulos,jpenedones,jonathan.campbell.toledo,baltvanrees,pedrogvieira\}}
\\ \vspace{1.2cm}
\footnotesize{\textit{
$^\text{\tiny 1}$Theoretical Physics Department, CERN, Geneva, Switzerland\\
$^\text{\tiny 2}$Institute of Physics, \'Ecole Polytechnique F\'ed\'erale de Lausanne, CH-1015 Lausanne,
Switzerland\\
$^\text{\tiny 3}$Centro de F\'isica do Porto,  Departamento de F\'isica e Astronomia,\\
Faculdade de Ci\^encias da Universidade do Porto,
Rua do Campo Alegre 687, 4169-007 Porto, Portugal\\
$^\text{\tiny 4}$Perimeter Institute for Theoretical Physics,
Waterloo, Ontario N2L 2Y5, Canada\\
$^\text{\tiny 5}$Centre for Particle Theory, Department of Mathematical Sciences, Durham University, Lower Mountjoy, Stockton Road, Durham, England, DH1 3LE\\
$^\text{\tiny 6}$ICTP South American Institute for Fundamental Research, IFT-UNESP, S\~ao Paulo, SP Brazil 01440-070}  
\vspace{4mm}
}

\par\vspace{1.5cm}

\textbf{Abstract}\vspace{2mm}
\end{center}
We consider constraints on the S-matrix of any gapped, Lorentz invariant quantum field theory in $1+1$ dimensions due to crossing symmetry and unitarity.   In this way we establish rigorous bounds on the cubic couplings of a given theory with a fixed mass spectrum. In special cases we identify interesting integrable theories saturating these bounds.  Our analytic bounds match precisely with numerical bounds obtained in a companion paper where we consider massive QFT in an AdS box and study boundary correlators using the technology of the conformal bootstrap.

\noindent

\setcounter{page}{1}
\renewcommand{\thefootnote}{\arabic{footnote}}
\setcounter{footnote}{0}

\setcounter{tocdepth}{2}

 \def\nref#1{{(\ref{#1})}}

\newpage

\tableofcontents

\parskip 5pt plus 1pt   \jot = 1.5ex

\newpage
\section{Introduction} 

The idea of constraining physical observables through a minimal set of indisputable principles is what is commonly referred to as \textit{bootstrap} philosophy. It shows up in various incarnations, the most well known being perhaps the integrable bootstrap, the conformal bootstrap and the S-matrix bootstrap. 

Since its inception \cite{Z77}, Zamolodchikovs' integrable bootstrap developed into an ironed out recipe for attacking various two dimensional theories, with non-linear sigma models and the Ising field theory as prototypical examples. It is often the only available analytic tool for studying such strongly coupled quantum field theories.  The conformal bootstrap works beautifully in two dimensions \cite{BPZ} where it allows for analytic description of a plethora of conformal field theories. In higher dimensions, the bootstrap had been dormant for decades until the seminal work \cite{RRTV}. This work gave rise to a new research field where one looks for bounds on the couplings and spectra of conformal field theories by exploiting crossing and reflection positivity. Using numerical algorithms, one rules out particular couplings or spectra by searching for linear functionals which yield impossibilities when acting on the crossing symmetry relations. Finally we have the very ambitious S-matrix bootstrap program -- which was very popular in the sixties, see e.g. \cite{books} for nice books on the subject -- which tries to completely determine S-matrix elements by exploring the analytic properties of these objects to its fullest.  With the development of efficient perturbative techniques and with the appearance of quantum chromodynamics, this program lost part of its original motivation and sort of faded away in its original form, morphing into string theory.\footnote{The formidable recent progress in our understanding of scattering amplitudes in gauge theories is a partial revival of this program, albeit for massless particles (the original S-matrix bootstrap was mostly aimed at the scattering of massive particles).  Also see \cite{Caron-Huot:2016icg, Sever:2017ylk} for some impressive recent progress in the S-matrix bootstrap of theories of weakly interacting higher spin particles.}

In this paper we observe amusing new connections between these various bootstrap branches: We will revisit the \textit{S-matrix bootstrap} for massive particles using a setup which is strongly inspired by the recent \textit{conformal bootstrap} bounds story and our results will make direct contact with both the \textit{integrable bootstrap} and the \textit{conformal bootstrap}. 

In a massive, strongly coupled quantum field theory the position of the poles of the S-matrix elements encode the mass spectrum of the theory while the magnitude of the residues measure the various interaction strengths, i.e. couplings. We will start a program aimed at carving out the space of massive quantum field theories by trying to establish upper bounds on couplings given a fixed spectrum of masses (of both fundamental particles and their bound states). The physical intuition motivating the existence of such bounds is that as couplings become larger the binding energy of any associated bound states increases -- that is, the bound state masses decrease and new bound-states may be pulled down from the continuum.   Thus it is reasonable to expect that for a fixed spectrum the coupling cannot be arbitrarily large.  


In this paper we systematically study these bounds in two dimensions where everything is simpler in the S-matrix world (the kinematical space simplifies significantly and crossing symmetry can be taken care of very explicitly). Not only do we find the above mentioned bounds but we also manage to identify known integrable theories which saturate the bounds at special points.  We hope these results will constitute the first steps in a general program aimed at extending the successful CFT bootstrap to massive QFT's.  


In a companion paper \cite{PaperA} we analyzed this problem from the conformal bootstrap point of view. There we put the  massive QFTs in an Anti de Sitter box. This induces conformal theories living at the AdS boundary which we can numerically study by means of the conformal bootstrap. The spectrum of dimensions and structure constants of these conformal theories can be translated back to the spectrum of masses and couplings of the quantum field theory in the bulk.  The analytic bounds described below by means of the S-matrix bootstrap turn out to beautifully match those from the conformal bootstrap numerics. This constitutes a non-trivial check both of the analytic results described here as well as the AdS construction proposed in \cite{PaperA} and the associated numerics. 


\section{Amplitude Bootstrap}  

Our main object of study will be the~$2\to 2$ S-matrix elements of a relativistic two dimensional quantum field theory. We will further focus on the elastic scattering process involving identical chargeless particles of mass $m$. For the most part, we shall take the external particles to be the lightest in the theory.\footnote{
Strictly speaking, what we shall use is that any two particle cut in the theory  opens up after the two particle cut of the external particles in this S-matrix element. 
The $2\to 2$ S-matrix element of the lightest particles is also free of Coleman-Thun singularities \cite{CT} (which render the analysis more involved and which will not be considered here). Sometimes, symmetry alone forbids such cuts or poles. In those case, the restriction to the lightest particle can be relaxed.  
}

\begin{figure}[t]
\begin{center}
\includegraphics[scale=0.6]{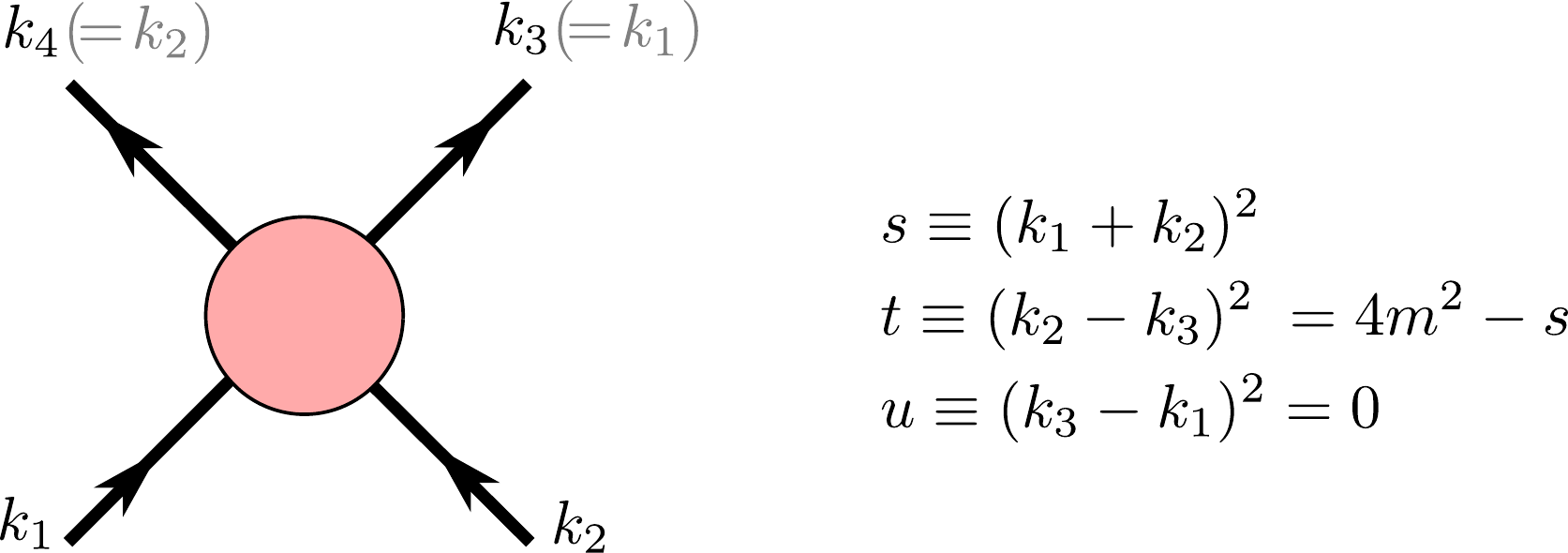} 
\end{center}
\caption{The $2\rightarrow 2$ S-matrix element.  Time runs vertically in this figure. In two dimensions energy-momentum conservation implies there is only one independent Mandelstam variable such that~$S=S(s)$ with $\sqrt{s}$ the centre of mass energy.}\label{SMatrix}
\end{figure}

Let us very briefly review a few important properties of this object, setting some notation along the way. 
A major kinematical simplification of $2\to 2$ scattering in two dimensions is that there is only a single independent Mandelstam invariant. In particular, for scattering involving particles of identical masses there is zero momentum transfer as depicted in figure~\ref{SMatrix}. 
If all external particles are identical, crossing symmetry which flips $t$ and $s$ simply translates into\footnote{Interchanging particles 3 and 4 leads to $t=0$, $u=4m^2-s$ and the same amplitude $S(s)$.}
\beq
S(s)=S(4m^2-s) \,, \la{crossing}
\eeq 
while unitarity states that for physical momenta, i.e for centre of mass energy greater than~$2m$, probability is conserved, 
\beq
|S(s)|^2 \le 1 \,, \qquad s>4m^2 \,.  \la{unitarity}
\eeq
We shall come back to this relation in more detail below, in section \ref{CDDsec}. 
\begin{figure}[t]
\begin{center}
\includegraphics[width=0.8\linewidth]{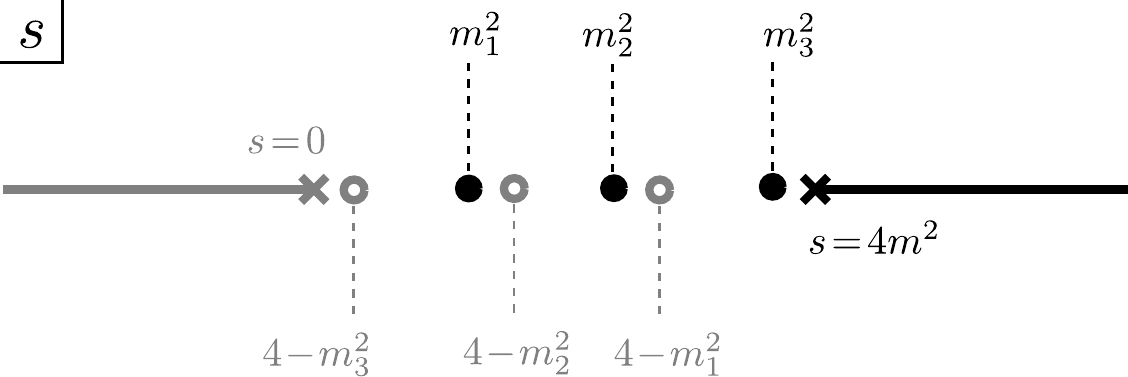} 
\end{center}
\caption{Analytic properties of the S-matrix element $S(s)$ for the scattering of the lightest particles of the theory. We have a cut starting at $s=4m^2$ corresponding to the two particle production threshold. As implied by (\ref{crossing}), we have another cut starting at $t=4m^2$ (or $s=0$) describing particle production in the  $t$-channel process. 
The segment $s\in[0,4m^2]$ between the two particle cuts is where most of the action takes place for us. It is here that poles corresponding to fundamental particles or their bound-states can appear as in (\ref{near pole of S}). We distinguish $s$ and $t$ channel poles (solid and empty circles respectively) by the sign of their residues. When the external particles are not the lightest in the theory, we sometimes have more singularities such as further two particle cuts and/or Coleman-Thun poles.
}\label{cutsS}
\end{figure}

Finally, we have the analytic properties of $S(s)$ depicted in figure \ref{cutsS}. Of particular importance for us are the S-matrix poles located between the two particle cuts. 
Such poles are associated to single-particle asymptotic states. Note that there is no conceptual difference between fundamental particles or bound-states here. We shall denote both as \textit{particles} in what follows. The poles in $S$ always come in pairs as
\beq\label{near pole of S}
S \simeq  -\mathcal{J}_j\,\frac{g_j^2}{s-m_j^2} \qquad \text{and} \qquad  S \simeq -\mathcal{J}_j \frac{g_j^2}{4m^{2}-s-m_j^2} \,, \qquad \Big( \,\mathcal{J}_j = \frac{m^{4}}{2 m_{j} \sqrt{4m^{2}-m_{j}^{2}}}\,\Big)
\eeq
corresponding to an $s$- or $t$-channel pole respectively. Here we normalize $g_j^2$ to be the residue in the invariant matrix element $T$ which differs from $S$ by the subtraction of the identity plus some simple Jacobians related to the normalization of delta functions in the connected versus disconnected components. This justifies the prefactors $\mathcal{J}_j$ in (\ref{near pole of S}).\footnote{
We have ${\bf S}\equiv {\bf1} \times S(s)= {\bf1}+ i (2\pi)^{2} \delta^{(2)}(P) \; T  $.  The contribution ${\bf 1} = (2\pi)^{2}4E_{1}E_{2} ( \delta(\vec{k}_{1}-\vec{k_{3}})\delta(\vec{k}_{2}-\vec{k_{4}})+(\vec{k}_{1} \leftrightarrow \vec{k_{2}}))$ represents the (disconnected) contribution of the free propagation while $T$ accounts for the connected contribution. Here we are we denoting the spatial momentum as $\vec{k}$ even though it is just a number just to distinguish it from the 2-momentum $k$.
Now, the delta function multiplying $T$ is the energy-momentum conservation delta function $\delta^{(2)}(P)=\delta^{(2)}(k_{1}+k_{2}-k_{3}-k_{4})$. On the support of the solution $\vec{k}_{1}=\vec{k_{3}}$, $\vec{k}_{2}=\vec{k_{4}}$ we have $4 E_{1}E_{2} ( \delta(\vec{k}_{1}-\vec{k_{3}})\delta(\vec{k}_{2}-\vec{k_{4}})+(\vec{k}_{1} \leftrightarrow \vec{k_{2}})) = 2 \sqrt{s}  \sqrt{s-4m^{2}} \,\delta^{(2)}(P)$. This Jacobian relating the $\delta$-functions results in the denominator in the definition of $\mathcal{J}_j$ in \eqref{near pole of S}. The $m^4$ numerator is just dimensional analysis: it is there so that $g_1$ is dimensionless. In other words, as defined, $g_1$ is the coupling measured in units of the external mass.}
Note that we can always clearly tell the difference between an $s$- or a $t$-channel pole: since in a unitary theory $g_j^2$ is positive, an $s$-channel pole has a negative residue (in $s$) while a $t$-channel pole has a positive residue. 

This concludes the lightning review of two dimensional scattering. 
We now have all the ingredients necessary to state the problem considered in this paper. As input we have a fixed spectrum of stable particles of masses $m_1<m_2<\dots<m_{N}$ which can show up as poles in $S(s)$. Note that by definition of stable asymptotic state (be it a bound-state or a fundamental particle) we have $m_j<2m$. Note also that $m_1$ might be equal to the mass $m$ of the external particle itself -- if the cubic coupling is non-vanishing -- or not -- such cubic coupling might be forbidden by a $\mathbb{Z}_2$ symmetry for instance.  The question we ask is then what is the maximum possible value of the coupling to the lightest exchanged particle (i.e $g_1$) compatible with such a spectrum,
\beq\label{question}
g_1^\text{max} \equiv \underset{\text{fixed } m_j}{\displaystyle\text{max}} g_1 = \, ?
\eeq
Physically, we expect the right hand side to be less than infinity. After all, as we increase the coupling to $m_1$ we expect this to generate an attractive force mediated by the particle $m_1$ between the two external masses. At some point, this force is such that new bound states are bound to show up, thus invalidating the spectrum we took as input. This should then set a bound on $g_1$. This question bears strong resemblance with very similar questions recently posed in the conformal bootstrap approach mentioned above. There also we can put upper bounds on the OPE structure constants given a fixed spectra of scaling dimensions \cite{Caracciolo:2009bx}.

We will approach this simple problem from two complementary angles. First in section \ref{DispSec} we will combine numerics with dispersion relation arguments to find a numerical answer. In section \ref{CDDsec} we present an analytic derivation of this bound exploring the power of analyticity and of two dimensional kinematics further.

\subsection{Dispersion Relations and the Numerical Bootstrap} \la{DispSec}

On the physical sheet the $S$ matrix has singularities corresponding to physical processes but is otherwise an analytic function.  Analyticity places strong constraints on $S(s)$ which can be summarized in a so-called dispersion relation which relates the $S$ matrix at any complex $s$ to its values at the cuts and poles, see e.g. \cite{books}.  
To set the notation and to specialize to two dimensions, we briefly recall the argument here. We start with the identity
\beq\label{cauchy}
S(s)-S_\infty = \oint\limits_{\gamma}\frac{dx}{2\pi i} \frac{S(x)-S_\infty}{x-s}  
\eeq
where $\gamma$ is a small counterclockwise contour around the point $s$ away from any pole or cut.  Now consider blowing the contour outward.  For simplicity we assume that $S(s)$ approaches a constant $S_\infty \in \[-1,1\]$ as $s\to \infty$ although this restriction can easily be lifted by means of so-called subtractions.\footnote{The basic idea of the subtraction procedure is to start with an identity of the form $
S(s)=\oint \frac{dx}{2\pi i} \frac{S(x)}{x-s}\prod_{a=1}^{n} \frac{s-x_{a}}{x-x_{a}}
$ where $n=1,2,\dots$ is the number of subtractions. As we blow up the contour, the integrand in the new identity is now more suppressed at large $x$ such that dropping the arc at infinity is safe for polynomially bounded amplitudes. In the end, this leads to similar albeit a bit more involved dispersion relations as compared to (\ref{dispersion}) below. We checked on a few examples that the numerics described below yield equivalent results with a few subtractions. More generally, assuming no essential singularity at $s=\infty$, we expect never to need more than $n=1$ in two dimensions. 
}
In this case we can drop the integration over the arcs at infinity so that we have only the integration around the poles and cuts giving
\beq\label{dispersion}
S(s) = S_\infty-\sum_{j} \mathcal{J}_j \(\frac{g_{j}^{2}}{s-m_{j}^{2}}+\frac{g_{j}^{2}}{4m^2-s-m_{j}^{2}}\)+\int\limits_{4m^{2}}^{\infty}dx \, \rho(x)\(\frac{1}{x-s}+\frac{1}{x-4m^2+s}\)
\eeq
where we have defined the discontinuity $2\pi i\, \rho(s)\equiv S(s+i 0)-S(s-i 0)$  
and we have further used the crossing equation \eqref{crossing} to replace the discontinuity across the t-channel cut in terms of the s-channel discontinuity.  

Equation \eqref{dispersion} is the sought after dispersion relation:  it simultaneously encodes the analyticity constraints as well as the crossing condition and thus provides a concrete framework for addressing the question \eqref{question}.  In this form, the question becomes:  what is the largest value of $g_{1}$ for which one can find $g_{2},...,g_{N}$ and $\rho(x)$ such that \eqref{unitarity} is satisfied?
\begin{figure}[t]
\begin{center}
\includegraphics[width=0.7\linewidth]{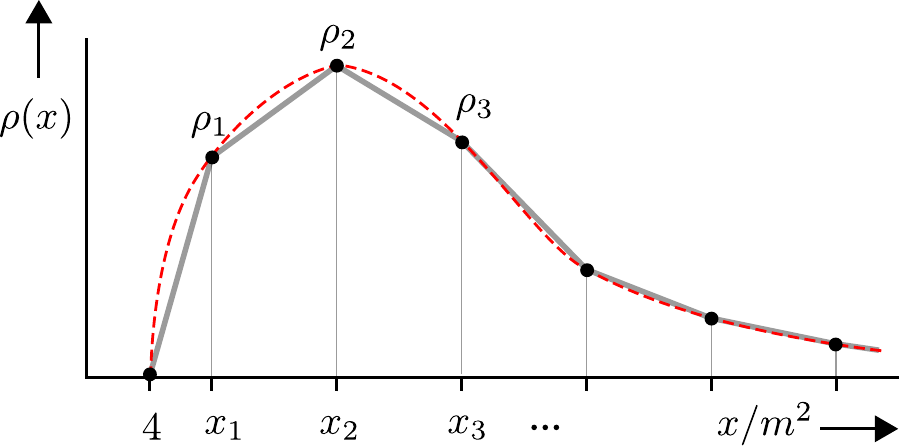} 
\end{center}
\caption[linear_spline]{Approximation of an arbitrary density with a linear spline.  The red dashed line represents some unknown $\rho(x)$ which we approximate with the grey spline passing through the points $(\rho_{n},x_{n})$.  Explicitly we have 
$\rho(x) \approx \rho_{n}\frac{(x-x_{n+1})}{(x_{n}-x_{n+1})} + \rho_{n+1}\frac{(x-x_{n})}{(x_{n+1}-x_{n})}$ for $x\in[x_{n},x_{n+1}]$.  We use this approximation up to some cutoff $x_{M}$ after which we assume the density decays as $\rho(x) \sim 1/x$. That is, we have $\rho(x) \approx \rho_{M} \, x_{M}/x$ for $x \ge x_{M}$ which allows us to explicitly integrate the tail from $x_{M}$ to $\infty$.}\label{linear_spline}
\end{figure}

Let us describe a concrete numerical approach to this question.  Denote by $\rho_{n}$ the value $\rho(x_{n})$ where $x_{n}\in[4m^2,\infty)$.  We can choose a set of $x_{n}$ and approximate $\rho(x)$ by a linear spline connecting the points $(x_{n},\rho_{n})$ as shown in figure \ref{linear_spline}.  We can then analytically perform the integral in \eqref{dispersion} to obtain
 \beq\label{discrete dispersion}
S(s) \approx S_\infty-\sum_{j} \mathcal{J}_j \(\frac{g_{j}^{2}}{s-m_{j}^{2}}+\frac{g_{j}^{2}}{4m^2-s-m_{j}^{2}}\)+ \sum_{a=1}^{M} \rho_{a} K_{a}(s) \eeq
where $K_a(s)$ are explicit functions of $s$ given in appendix \ref{numerics}.  Evaluating this expression at some value $s_{0}>4m^2$ and plugging it into equation~\eqref{unitarity} gives us a quadratic constraint in the space of variables $g_j^2$, $\rho_n$ and $S_\infty$.  
 The space of solutions of the constraints is then the intersection of all these regions for all values of $s_{0}>4m^2$.\footnote{We can visualize this region as the intersection of many cylinders, given by equation  \eqref{discrete unitarity}, in a high dimensional space.} It now suffices to start inside this region and move in the direction of increasing $g_1^2$ until we hit the boundary of the region and can move no more.
 


In practice, these numerics are simple enough that they can be performed in a few seconds in \texttt{Mathematica} using the built-in function \texttt{FindMaximum} which allows one to search for the maximum value of a function inside of some constraint region. For more details see appendix~\ref{numerics}. 

\begin{figure}[t]
\begin{center}
\includegraphics[scale=1.5]{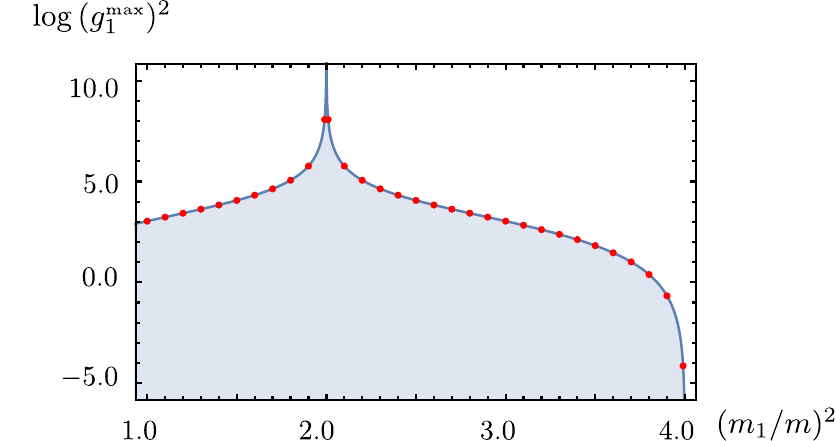} 
\end{center}
\caption{Maximum cubic coupling $g_1^\text{max}$ between the two external particles of mass $m$ and the exchanged particle of mass $m_1$. Here we consider the simplest possible spectrum where a single particle of mass $m_1$ shows up in the elastic S-matrix element describing the scattering process of two mass $m$ particles. The red dots are the numerical results. The solid line is an analytic curved guessed above (\ref{SGmatrix}) and derived in the next section. The blue (white) region corresponds to allowed (excluded) QFT's for this simple spectrum. }\label{gvsm1}
\end{figure}

\begin{figure}[t]
\begin{center}
\includegraphics[width=1\linewidth]{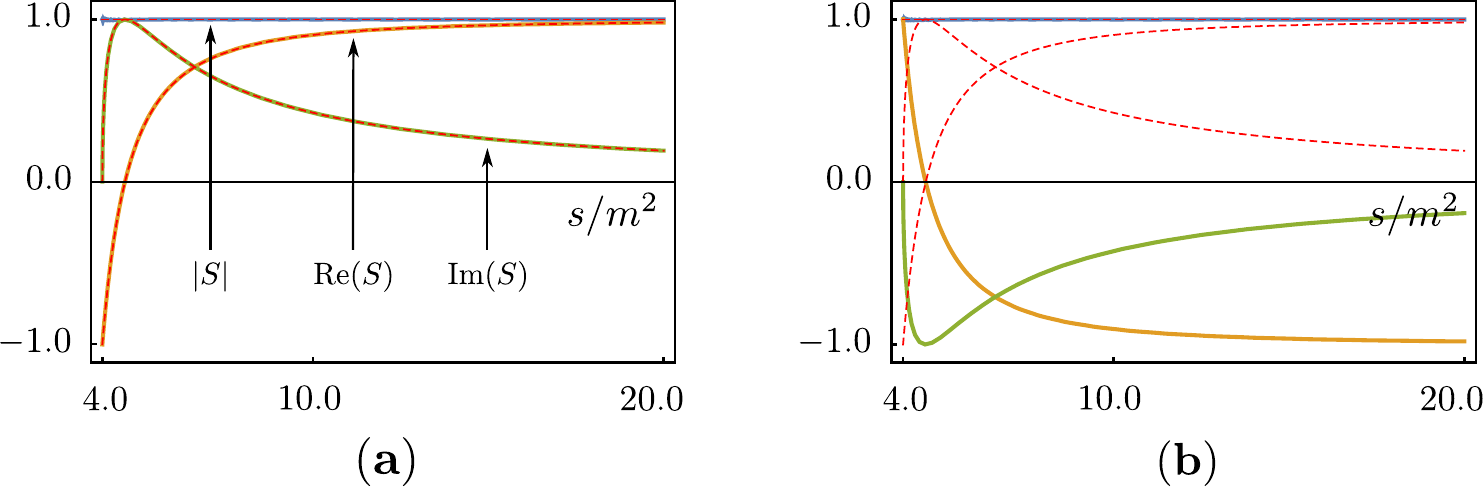} 
\end{center}
\caption{Result of numerics for (a) $m_1 = \sqrt{3}$ and (b) $m_1=1$.  In both figures the green, orange and blue curves are $\text{Im}(S)$, $\text{Re}(S)$, $|S|$ respectively.  Note that the blue curve is flat and equal to~$1$. In other words, the S-matrix that maximizes $g_1$ saturates unitarity at all values of $s>4m^2$. The red dashed lines are real part, imaginary part and magnitude of the sine-Gordon S-matrix \eqref{SGmatrix}.  In figure (a) the numerical results match perfectly with  \eqref{SGmatrix}, while in figure (b) the numerics give precisely $(-1)$ times the sine-Gordon S-matrix as explained in the text.} \label{SnumSCDD}
\end{figure}

To illustrate, consider the simplest possible example in which only a particle of mass $m_{1}$ couples to the external particle of mass $m$. In other words, we consider an S-matrix with a single s-channel pole whose residue we are trying to maximize.  We can then follow the procedure outlined above to find the maximum value of the coupling $g^\text{max}_{1}$ for each value of~$m_{1}/m$. The results are depicted in figures~\ref{gvsm1} and~\ref{SnumSCDD} . 

The numerical results depicted in these plots reveal various interesting features. First, we have the spike in  figure \ref{gvsm1}. It has a simple kinematical explanation. As $m_1 \to \sqrt{2} m$ the $s$- and $t$-channel poles in (\ref{dispersion}) collide and thus annihilate each other. As such we can no longer bound the residue at this point. The symmetry~$g_{1}^\text{max}(m_{1}^{2})=g_{1}^\text{max}(4 m^{2}-m_{1}^2)$ observed in the numerics is equally simple to understand. Each solution to the problem with $m_{1}>\sqrt{2} m$ can be turned into a solution to the problem with $m_{1}<\sqrt{2}m$ provided we re-interpret who is the $s$- and who is the $t$- channel pole which we can easily do if we  multiply the full S-matrix by $-1$. The plots in figures \ref{SnumSCDD} corroborate this viewpoint.  

Another interesting regime is that where the exchanged particle is a weakly coupled bound-state of the external particles, that is $m_1 \simeq 2m$. As $m_1 \to 2m$ we see in the numerics that the maximum coupling vanishes. This is an intuitive result:  only a small coupling can be compatible with this spectrum as a larger coupling would decrease the mass of the bound state.  Note that this corner of our bounds can be studied using perturbation theory \cite{Dashen:1975hd}.


Finally, and most importantly, we observe in the plots in figure \ref{SnumSCDD} that the numerical solutions for the $S$-matrices with the maximal residues actually saturate unitarity at \textit{all} values of~$s>4m^2$. This observation has immediate implications. It implies the absence of~$2\to n$ particle production for any $n>3$. After all, 
\beq
|S_{2\to 2}(s)|^2=1-\sum_\text{other stuff X} |S_{2\to X }(s)|^2 \, , \qquad s>4m^2 \,. \la{unitarityComplete}
\eeq
Absence of particle production is the landmark of integrable models. S-matrices which saturate unitarity often show up in the integrable bootstrap and can usually be determined analytically. When $m_{1}>\sqrt{2}m$, for instance, there is a well known S-matrix obeying~$|S(s)|^2=1$ for $s>4m^2$ and with a single bound-state $s$-channel pole at $s=m_{1}^2$. It is the Sine-Gordon S-matrix describing the scattering of the lightest breathers in this theory; and the bound state is the next-to-lightest breather. Explicitly, it reads \cite{Z76, Arefeva:1974bk}
\beq
S_{SG}(s)=\frac{\sqrt{s} \sqrt{4 m^2-s}+m_1\sqrt{4 m^2-m_1^2} }{\sqrt{s} \sqrt{4 m^2-s}-m_1 \sqrt{4 m^2-m_1^2}} \,. \la{SGmatrix}
\eeq
The dashed lines in figure \ref{SnumSCDD}a correspond to the values of the real and imaginary parts of this analytic S-matrix. Clearly, it agrees perfectly with the numerics. Our claim is that there is no unitary relativistic quantum field theory in two dimensions whose S-matrix element for identical particles has a single bound-state pole at $s=m_{1}^2>2m^2$ and a bigger residue than that of the Sine-Gordon breather S-matrix.

Also, according to what we discussed above, we conclude (and cross-check in figure \ref{SnumSCDD}b) that the S-matrix with the maximum coupling $g_1^\text{max}$ and with a bound-state $m_{1}<\sqrt{2}m$ is given by an S-matrix which differs from the Sine-Gorgon S-matrix by a mere minus sign, $S(s)=-S_{SG}(s)$. We do not know of any theory with this S-matrix.\footnote{If you do and would drop us an e-mail that would be greatly appreciated. It is also conceivable that such a theory does not exist at all. The bound for $m_1/m>\sqrt{2}$ must be optimal since Sine-Gordon theory exists. However, the left region of the plot in figure \ref{gvsm1} for $m_1/m<\sqrt{2}$ might still move down as we include into the game further constraints such as those coming from S-matrix elements involving other particles in the theory as external states. This is analogous to what has been done in the conformal bootstrap \cite{Kos:2014bka}.}

In the next section we will explain that the phenomenon we encountered empirically here -- i.e. saturation of unitarity -- is actually generic and not merely a peculiarity of this simplest example with a single exchanged particle. This will open the door toward an analytic derivation of $g_1^\text{max}$ for any bound-state mass spectrum of $\{m_1/m,m_2/m,\dots\}$. 

\subsection{Castillejo-Dalitz-Dyson factors and the Analytic Bootstrap} \la{CDDsec}
An important hint arose from the numerics of the last section: for the simplest possible mass spectrum (with a single s-channel pole), we found that the optimal S-matrix --  leading to a maximum coupling~$g_1^\text{max}$  -- \textit{saturates unitarity} at any $s>4m^2$ (see the blue curves in figure~\ref{SnumSCDD}). This simple example suggests that one should be able to borrow standard machinery from the integrable bootstrap literature to tackle this problem analytically. This is what we pursue in this section. Ultimately, this will lead to an analytic prediction for $g_1^\text{max}(m_1/m,\dots)$ for an arbitrary spectrum of masses.  Actually, our analysis will determine the full S-matrix element corresponding to this maximal coupling.

\begin{figure}[t]
\begin{center}
\includegraphics[scale=1]{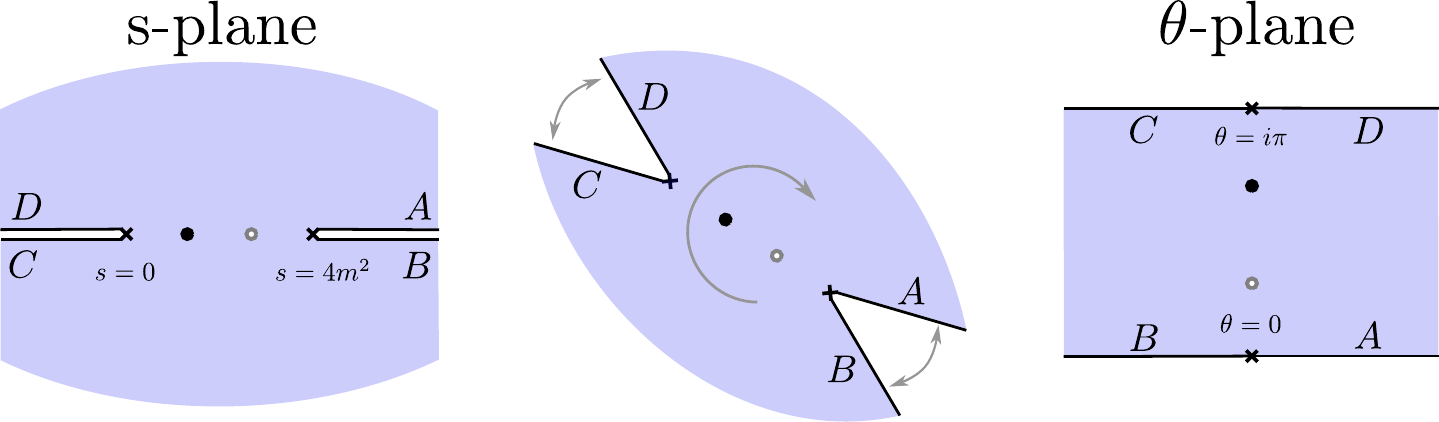} 
\end{center}
\caption{Mapping from $s$ to $\th$.  The map ``opens'' the cuts and rotates clockwise by $\pi/2$.  The physical sheet of the s-plane is mapped to the strip $\text{Im}(\theta)\in [0,\pi]$ with $s=0$ $(s=4m^{2})$ mapping to $\th=i\pi$ ($\th=0$).
} \label{theta_plane} \end{figure}

To proceed, it is convenient to change variables from $s$ to the usual hyperbolic rapidity~$\theta$ with $s=4m^2 \cosh^2(\theta/2)$. The mapping from $s$ to $\th$ is shown in figure \ref{theta_plane}. The strip $\text{Im}(\theta)\in [0,\pi]$ covers the full physical $s$-plane of figure \ref{cutsS} and is thus called the physical strip. 
We recall in appendix \ref{thetaAp} a few useful properties of this parametrization. 
In terms of $\theta$ we write crossing and unitarity as 
\beqa
S(\theta)=S(i\pi-\theta) \, , \qquad  S(\theta+i0)S(-\theta+i0) = f(\theta)  \,, \la{eps}
\eeqa
Where $f$ is the right hand side of (\ref{unitarityComplete}) which we do not know.  We do know that, by definition, this absorption factor takes values in $ f \in [0,1]$ for physical momenta, that is for~$\theta \in \mathbb{R}$. Now, a solution to (\ref{eps}) can always be written as
\beq
S(\theta)=S_\text{CDD}(\theta) \exp\( -\int\limits_{-\infty}^{+\infty} \frac{d\theta'}{2\pi i} \frac{\log f(\theta')}{\sinh(\theta-\theta'+i0)}  \)
\eeq
where the exponential factor is a particular solution to (\ref{eps}) -- known as the minimal solution -- while $S_\text{CDD}(\theta)$ is a solution to (\ref{eps}) with~$f=1$. Note that the minimal solution has no poles (or zeros) in the physical strip; any poles (or zeros) are taken into account by $S_\text{CDD}$. 


It is now rather straightforward to understand why the process of maximizing the coupling to the lightest exchanged particle leads to S-matrices which saturate unitarity, i.e. for which~$f=1$. Indeed, using the fact that $f$ is an even function, we can symmetrize the integral in the minimal solution to get 
\beq
S(i t)=S_\text{CDD}(i t) \times \exp\Big( \int_{-\infty}^{+\infty} \frac{d\theta'}{2\pi } \underbrace{ \frac{\sin(t) \cosh(\theta')}{|\sinh(it-\theta')|^2}}_{\text{positive for $t\in [0,\pi]$}}   \times \underbrace{\log f(\theta')}_{\text{negative} } \Big)\,.
\eeq
in the segment $t\in [0,\pi]$ corresponding to $s\in [0,4m^2]$ where the potential poles of the S-matrix lie. We see that the minimal solution always decreases the magnitude of the S-matrix in this segment unless $f=1$. Therefore, if we are to maximize some residue in this region it is always optimal to set $f=1$. 
This simple observation explains  the saturation of unitarity observed experimentally in the last section and establishes it for any spectrum of poles. 

\begin{figure}[t]
\begin{center}
\includegraphics[scale=1.0]{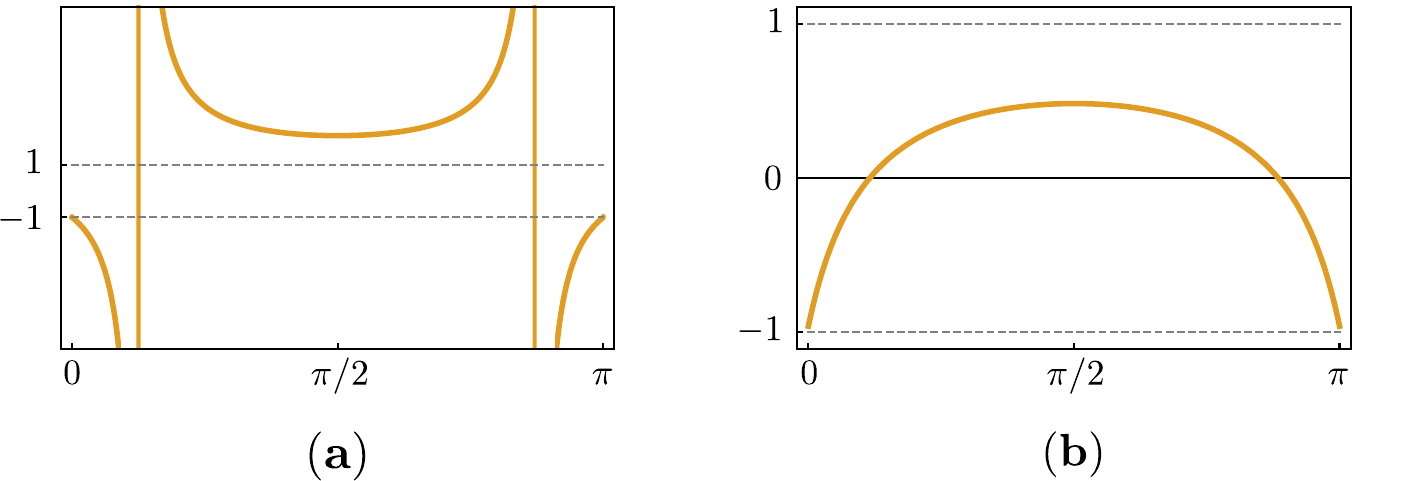} 
\end{center}
\caption{Panel (a) shows a CDD pole $[\pi/8]$ for $\theta$ purely imaginary between $0$ and $i\pi$.  Note that the magnitude of this factor is always greater or equal $1$.  Also note that it is positive between its $s$- and $t$- channel poles, while the tails of the function are negative.  Panel (b) shows a CDD zero $[-\pi/8]$ in the same interval.  The magnitude of this function in this interval is always less than or equal to $1$.   
} \label{CDD_pole_and_zero}
\end{figure}

Next we have the Castillejo-Dalitz-Dyson (CDD) term which solves the homogenous problem
\beqa
S_\text{CDD}(\theta)=S_\text{CDD}(i\pi-\theta) \, , \qquad  S_\text{CDD}(\theta)S_\text{CDD}(-\theta) = 1 \,. \la{eps2}
\eeqa
There are infinitely many solutions to this homogenous problem which we can construct by multiplying any number of so-called CDD factors \cite{Mussardo:1999aj}, 
\beqa\label{general CDD sol}
S_\text{CDD}(\theta)=\pm \prod_{j} [\alpha_j] \,, \qquad [\alpha]\equiv\frac{\sinh(\theta)+i \sin(\alpha)}{\sinh(\theta)-i \sin(\alpha)} \,.
\eeqa
Without loss of generality, we take $\alpha$ to be in the strip $\rm{Re}(\alpha)\in [-\pi,\pi]$. Still, depending on its value these CDD factors $\[\alpha\]$ can represent very different physics. There are basically three different instances to consider:

Consider first the case when $\alpha$ is in the right half of the above mentioned strip, i.e.  $\rm{Re}(\alpha)\in [0,\pi]$. In this case the corresponding CDD factor will have a pole at $\theta=i\alpha$ in the physical strip. Because of locality such poles should always be located in the segment $s\in [0,4m^2]$ corresponding to $\theta$ purely imaginary between $0$ and $i\pi$. Therefore if $\alpha$ is in the right half of its strip, it ought to be purely real with $\alpha \in[0,\pi]$. In this case, the CDD factor $\[\alpha\]$ is referred to as a {\bf \underline{CDD-pole}}; an example is plotted in figure \ref{CDD_pole_and_zero}a. Clearly, these factors are very important. They are the only factors which give rise to poles in the $S$-matrix corresponding to stable asymptotic particles. 

\begin{figure}[t]
\begin{center}
\includegraphics[scale=1]{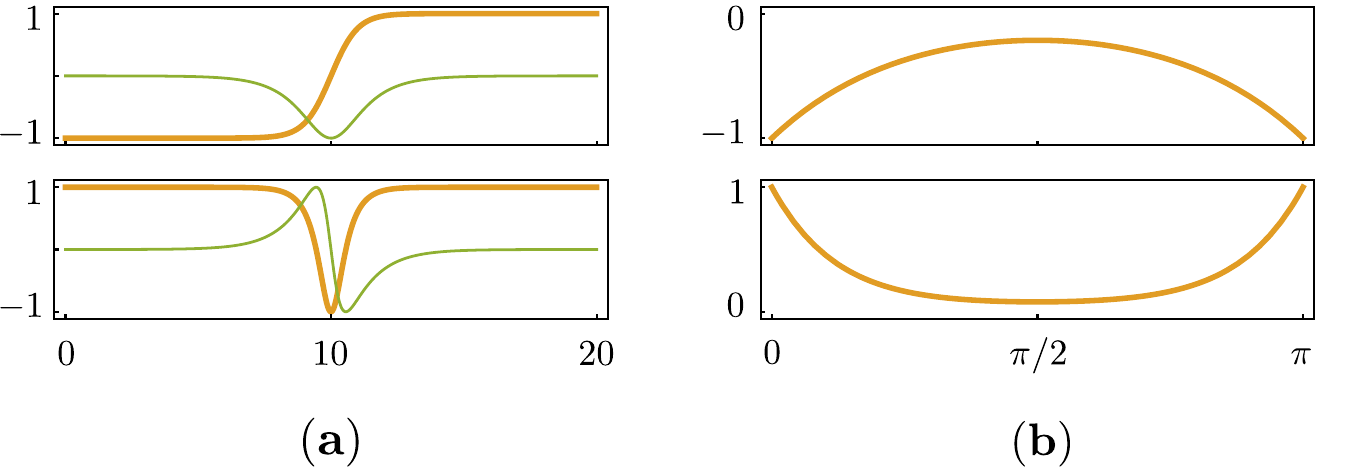} 
\end{center}
\caption{Panel (a) shows the behaviour of two types of CDD resonances for real $\th$.  The upper and lower plots show $[-\pi/2+10 i]$ and $[-\pi/5-10 i] [-\pi/5+10 i]$ respectively.  The thick orange curve is the real part while the thin green curve is the imaginary part.  Resonances can be added at very little cost. If some parameters are large, for example, their effect only shows up at very high energies nearly not affecting low energy physics.  Panel (b) shows the behaviour of two resonance factors for $\theta$ purely imaginary between $0$ and $i\pi$.  The upper and lower panel show $[-\pi/2+i]$ and $[-\pi/3-i][-\pi/3+i]$ respectively.  In the former case the resonance factor is purely real in this interval while in the later case the product is real although the individual factors are not. Note that in this interval CDD resonances always have magnitude less than $1$ and that each individual CDD resonances never changes sign.} \label{resonancesFig}
\end{figure}

When $\alpha$ is in the left half of the above mentioned strip there are less physical constraints on its admissible values. The reason is that in this case the corresponding factor induces a pole at $\theta=i\alpha$ which is now no longer in the physical strip. In terms of $s$ it would be on another sheet after crossing some of the cuts in figure \ref{cutsS}. A priori, there is not much we can say about possible positions of poles which leave the physical strip. It is still convenient to separately consider two possible cases. If $\alpha$ is purely real in the left strip --  that is if $\alpha \in [-\pi,0]$ -- we say $\[\alpha\]$ is a {\bf\underline{CDD-zero}}. The reason is clear: such a factor has a zero at $\theta=-i\alpha$ inside the physical strip and along the very same segment where possible poles will be.  An example of a CDD zero is plotted in figure \ref{CDD_pole_and_zero}b.  We can also have complex values of $\alpha$ provided they are carefully chosen not to spoil real-analyticity of S-matrix which requires that $S(\theta)$ should be real in the segment between $0$ and $i\pi$. One possibility for example would be to have $\alpha=-\pi/2+i \beta$ where $\beta$ is purely real. Another option would be to have a pair of complex conjugate $\alpha$'s such that their product would lead to a real contribution in the above mentioned segment. Such CDD contributions also lead to zeros in the physical strips, this time at complex values of $\theta$. We refer to such factors as~{\bf\underline{CDD-resonances}}.  Examples of CDD resonances are plotted in figure \ref{resonancesFig}.

Let us now discuss some general features of these three CDD factors which are relevant for our  purposes.  We see in figure \ref{CDD_pole_and_zero}a that a CDD-pole factor has magnitude greater than one at any point in the segment $\theta=[0,i\pi]$.  On the other hand from figure \ref{CDD_pole_and_zero}b and \ref{resonancesFig} we see that CDD-zeros and CDD-resonances have magnitude always smaller or equal to one in this segment.  As such, one may (incorrectly) conclude that the S-matrix which maximizes $g_1$ and is compatible with a given spectrum of asymptotic stable particles~$\{m_1/m,m_2/m,\dots\}$ is simply given by a product of CDD-poles, one for each stable particle. 

This is too hasty for the simple reason that such a naive product of CDD-poles will generically have wrong signs for the corresponding residues contradicting \eqref{near pole of S}.\footnote{Translating \eqref{near pole of S} to $\theta$-space we have that a proper s-channel pole corresponding to a mass~$m_j^2=4\cosh^2(\theta_j/2)$ should behave as $S\simeq i\Gamma_j^2/(\theta-\theta_j)$ with $\Gamma_j^2$ positive and related to $g_j^2$ by some simple Jacobians. Correspondingly, the associated t-channel pole will be located at $\theta=i\pi-\theta_j$ and will have a negative residue $S\simeq -i\Gamma_j^2/(\theta-i\pi+\theta_j)$.} Hence, a more thoughtful conclusion is that while we can indeed discard any CDD-resonances, CDD-zeros are sometimes necessary. In contradistinction with the CDD-resonances and also with the minimal solution discussed above, CDD-zeros change sign in the segment $\theta=[0,i\pi]$ so we can -- and must -- use them to flip the wrong signs of any residues. The correct prescription is therefore to dress the product of CDD-poles by a potential overall sign plus a minimal amount of CDD-zeros such that the signs of all the residues come out right. The position of the CDD-zeros is then fixed such that $g_1$ is maximal.  Appendix \ref{CDDgeneral} contains the final outcome of this maximization problem for the most general mass spectrum.  Rather than give a derivation of this general result, we find it is more useful to consider a few simple examples from which the general result follows as a natural extrapolation.   To this end in the next section we work out a few illustrative examples in full gory detail.

\subsection{Analytic Bootstrap Examples}
Let us begin with the simplest case in which there is a single particle with $m_{1} < 2m$.  We wish to maximize the coupling for the process $m+m \to m_{1}$.  This was the case considered in section \ref{DispSec} and for which the results of the numerics are given in figures \ref{SnumSCDD} and \ref{gvsm1}.  Since there is only a single bound state, we require only one pole and thus the solution is given by $S=\pm [\a_{1}]$ where $\a_{1}$ is fixed by the condition 
\beq
m_{j}^{2}=4 \cosh^{2}(i \a_{j}/2)
\eeq
and the $\pm$ is fixed such that the residue of the $s$-channel pole is positive.   This leads to $S=[\a_{1}]$ for $m_{1}>\sqrt{2}$ and $S=-[\a_{1}]$ for $m_{1}<\sqrt{2}$.

\begin{figure}[t]
\begin{center}
\includegraphics[scale=1]{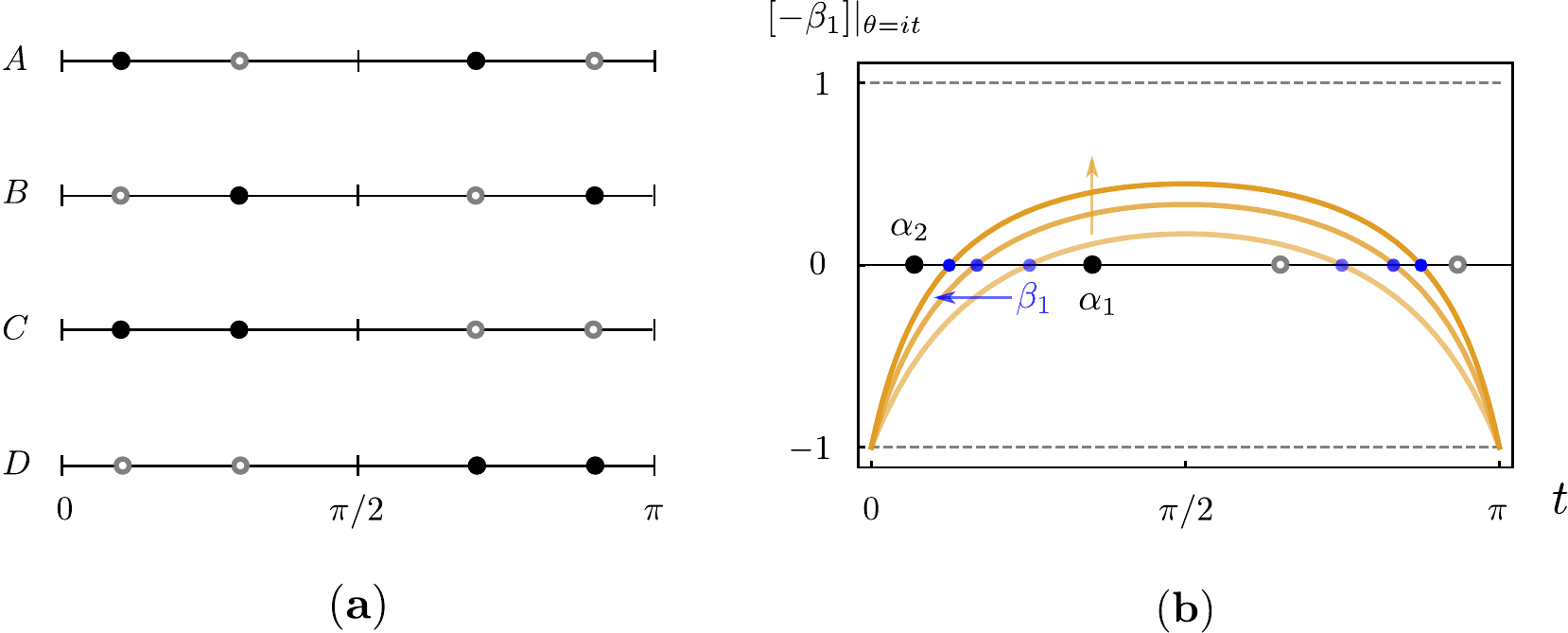} 
\end{center}
\caption{Panel ({\bf a}) shows the four possible configurations of poles for a spectrum $m_{1}<m_{2}<2m$ and no cubic coupling.  Cases $A$ and $B$ correspond to $m_{1}<\sqrt{2}<m_{2}$ the former with $m_{1}^{2}>4-m_{2}^{2}$ and the latter with $m_{1}^{2}<4-m_{2}^{2}$.   Cases $C$ and $D$ correspond to $\sqrt{2}<m_{1}<m_{2}$ and $m_{1}<m_{2}<\sqrt{2}$ respectively.  The residues of a product of CDD factors alternate between positive and negative since a CDD factor changes sign at each of its poles and nowhere else.  Thus in case $A$ and $B$ we can arrange for \eqref{near pole of S} to be satisfied simply by fixing the overall sign of the S-matrix.  Cases $C$ and $D$ cannot be repaired in this way.  Instead we must multiply by a CDD zero in order to fix the signs.  Panel ({\bf b}) shows a CDD zero factor $[-\beta_{1}]$ with $\a_{2}<\beta_{1}<\a_{1}$ such that is changes sign between the two s-channel poles and also between the two t-channel poles.  In this way the product $\pm [\a_{1}][-\beta_{1}][\a_{2}]$ will have the correct residues (the overall sign  can be then fixed as in cases $A$ and $B$).  The precise value of $\beta_{1}$ must then be fixed to maximize $g^{2}_{1}$ which is the residue at $\a_{1}$.  We see that $[-\beta_{1}]$ grows monotonically as we shift the zero to the left toward $\a_{2}$.  Optimizing then implies that we must collide this zero with the pole at $\a_{2}$. } \label{optimizing_zero}
\end{figure}

Now suppose we have two particles such that $m_{1}<m_{2}<2m$ and again we wish to maximize the coupling for the process $m+m \to m_{1}$.  Clearly we should start with at least two CDD factors to accommodate bound-state poles at $s=m_{1}^{2}$ and $s=m_{2}^{2}$.  However, the analysis is complicated by the requirement that the residues of these poles be positive since each individual CDD factor changes sign at its poles (see figure \ref{CDD_pole_and_zero}a).  We must consider the four distinct configurations of s- and t-channel poles shown in figure \ref{optimizing_zero}a.
First consider cases $A$ and $B$  which correspond to $m_{1}<\sqrt{2}<m_{2}$.  Here the solution is simply given by $S=\pm\[\a_{1}\]\[\a_{2}\]$.  Once the correct overall sign is selected, the sign of the residues of the poles work out since the poles alternate between $s$ and $t$ channel.\footnote{We fix the overall sign as follows.  Notice from figure \ref{CDD_pole_and_zero}a that an individual CDD factor is positive between its poles and negative before and after -- i.e. the tails of the CDD factors are always negative.  Further, the pole of an individual CDD factor closest to $i\pi$ has the form $ i \, (-1) \times (\text{positive})$.  Thus, for a general product of such factors the sign of the residue closest to $i\pi$ has the form $ i \, (-1)^{N} \times (\text{positive})$. If  $m_{1}^{2}>4-m_{2}^{2}$ this pole will be t-channel as in case $A$ of figure \ref{optimizing_zero}a and since $N=2$ we should choose the overall sign $(-1)$.  On the other hand when $m_{1}^{2}<4-m_{2}^{2}$ the first pole is s-channel is in case $B$ and thus we should choose the overall sign $(+1)$.  In general, configurations of poles which are related by reflection about $\pi/2$ have an S-matrix related by an overall sign.}

\begin{figure}[t]
\begin{center}
\includegraphics[scale=0.6]{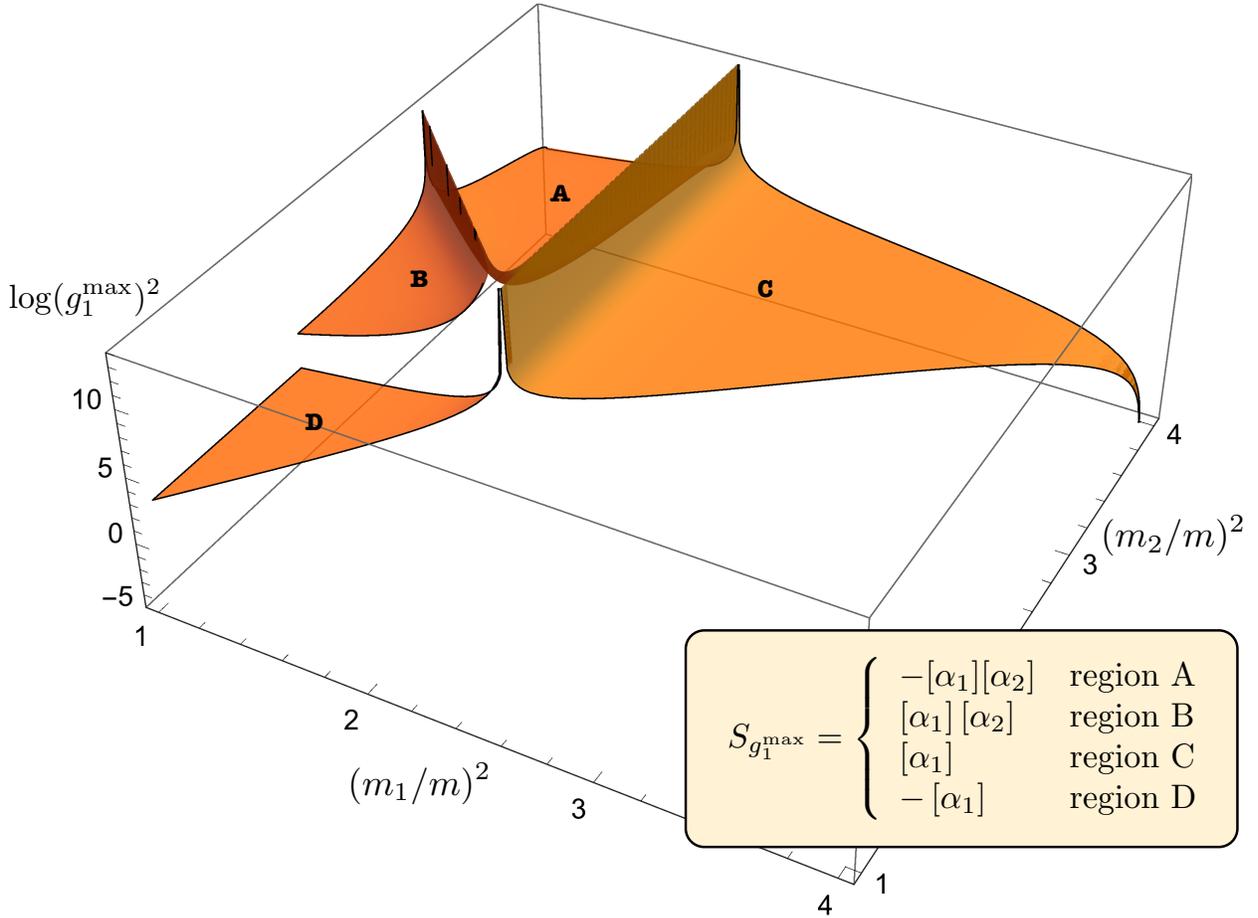} 
\end{center}
\caption{Maximal coupling $g_{1}^{\text{max}}(m_1/m,m_2/m)$ for the spectrum $m_{1}<m_{2}<2m$ and no cubic coupling.  Each region corresponds to one of the four configurations of poles shown in figure \ref{optimizing_zero}a.} \label{TwoMasses3}
\end{figure}

Now consider the case $C$ in figure \ref{optimizing_zero}a which corresponds to $\sqrt{2}<m_{1}<m_{2}$.  Now a simple product of two CDD poles cannot have the correct signs for its residues. The signs alternate at each pole but we have two consecutive s-channel poles with no t-channel pole in between.  To correct for this, we are forced to insert a CDD zero $[-\beta_{1}]$ between the two $s$-channel poles $\a_{2}<\beta_{1}<\a_{1}$.  Such a factor also has a zero between the two t-channel poles since it is crossing symmetric. The precise position of this zero is then fixed by the condition that $g_{1}^{2}$ be maximized -- i.e we want to maximize the value of the CDD zero at the position $\a_{1}$.   From figure \ref{optimizing_zero}b we see that this means we should move the zero as far away from $\a_{1}$ as possible.  In particular, it implies that we should collide the zero with the pole at $\a_{2}$, thus decoupling that state from the scattering of the lightest particle.  In other words, the the optimal $S$-matrix is given by $S=[\a_{1}]$ for $\sqrt{2}<m_{1}<m_{2}$.  Note that this does not contradict our assumption that there is a particle $m_{2}$ in the spectrum.  Rather, it simply implies that the $S$-matrix that maximizes $g_{1}$ has no coupling to this asymptotic state (i.e. $g_{2}=0$).   Lastly, case $D$ is related to $C$ by reflection about $\pi/2$ so in that case we have $S=-[\a_{1}]$.  The final result of all this analysis is summarized in figure \ref{TwoMasses3}.

\begin{figure}[t]
\begin{center}
\includegraphics[scale=0.6]{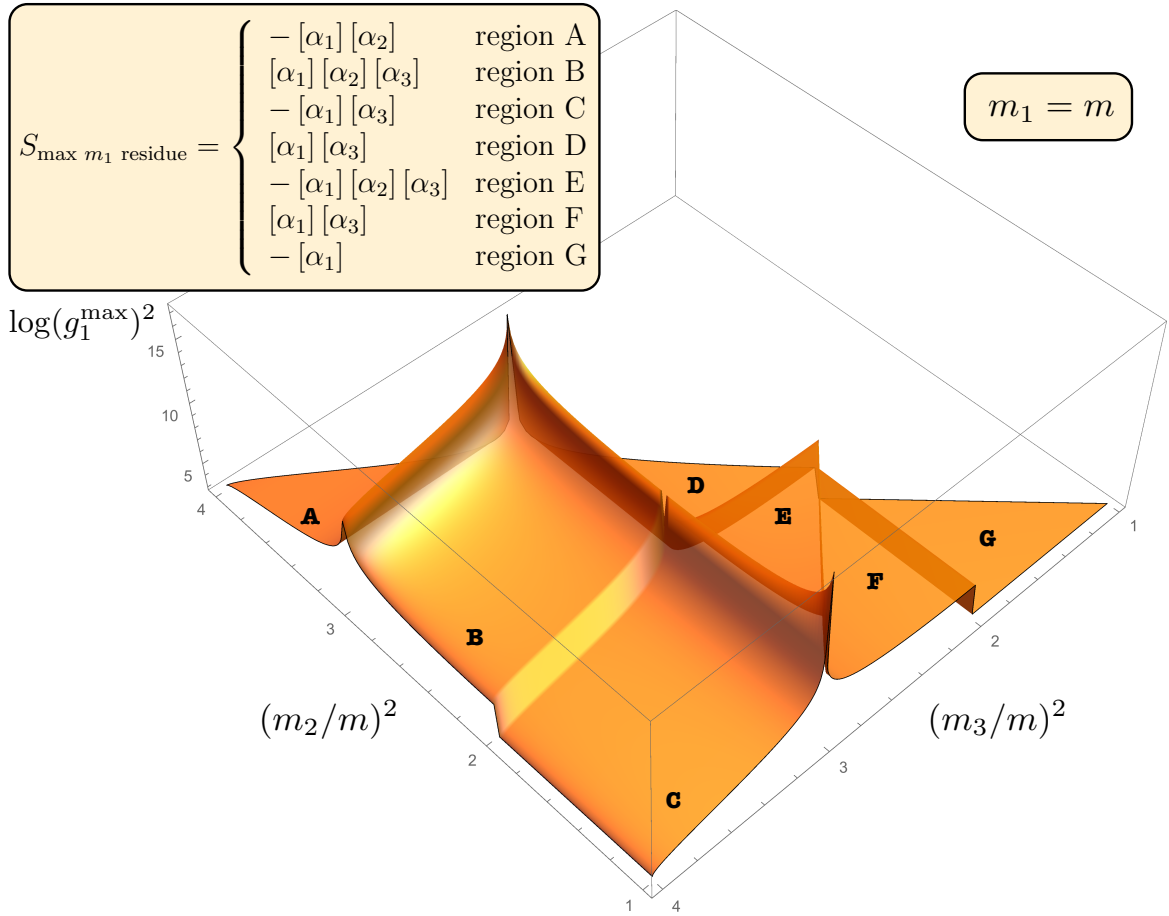} 
\end{center}
\caption{Maximal coupling $g_{1}^{\text{max}}(m_2/m,m_3/m)$ for the spectrum $m_{1}=m$ (i.e. a cubic coupling $m+m \to m$) and generic $m<m_{2}<m_{3}<2m$.} \label{ThreeMasses2}
\end{figure}

The case $N=2$ that we have just discussed demonstrates all the salient features of the general case.  In particular for a set of masses $m_{1}\!<\!m_{2}\!<\!...\!<\!m_{N}\!<\!2m$ corresponding to $\{\a_{1},...,\a_{N}\}$ the optimizing $S(s)$ will always be given by \eqref{general CDD sol} where the product runs over a {\it subset} of the masses.    The product is only over a subset because the collision of zeros and poles we observed in the $N=2$ case is a feature present in the general solution.  That is, whenever the poles do not alternate between s- and t-channel, we are forced to insert CDD zeros so that the residues obey \eqref{near pole of S}.  Maximizing with respect to the position of these zeros always forces them to collide with a pole, thus decoupling that state from the scattering process.  Precisely which poles get canceled is explained in appendix \ref{CDDgeneral}.  Finally, the overall sign in \eqref{general CDD sol} is fixed by considering whether the pole closest to $i\pi$ is s- or t-channel.  The end result of this analysis is formula \eqref{optimal} given in appendix \ref{CDDgeneral}.  As an application which will be relevant in the next section, in figure \ref{ThreeMasses2} we present the maximal coupling for the case $m_{1}=m$ (i.e. a cubic coupling $m+m \to m$) and generic $m_2$, $m_3$ satisfying $m<m_{2}<m_{3}<2m$.  Finally, we have verified in all these cases that performing the numerics of section \ref{DispSec} for the various configuration of poles confirms the CDD solutions given above.

We will now conclude with some comments regarding the CDD solution \eqref{general CDD sol}.  First we note 
this solution \eqref{general CDD sol} does not cover the full space of solutions of \eqref{eps2}.  If we allow for an essential singularity at $s=\infty$, then we can multiply \eqref{general CDD sol} by 
\beq
S_\text{grav} (s)= e^{i l_s^2 \sqrt{s(s-4)}} = e^{2 i l_s^2 m^2 \sinh \theta} 
\eeq 
with an arbitrary parameter $l_{s}^{2}$.  This solution, called a ``gravitational dressing factor''  was recently introduced in \cite{Dubovsky:2013ira}.  For our purposes we can rule out the possibility of such a factor since $S_{\text{grav}}\in[0,1]$ in the segment $\theta=[0,i \pi]$ and thus will always decrease the value of $g_{1}$.  We do not know any other solutions of \eqref{eps2} that could be used to increase the value of $g_{1}$. 

Second, note that the general CDD solution \eqref{general CDD sol} saturates unitarity ($|S_{\text{CDD}}|=1$ for $\th$ real) which implies the absence of particle production in the scattering $m+m$.  As we have already mentioned in section \ref{DispSec} absence of particle production is an indication of integrability.  Thus one may wonder if each point on the surfaces of figures \ref{TwoMasses3} and \ref{ThreeMasses2} correspond to some integrable model.  As we shall see in the next section, generic points in these plots can \textit{not} correspond to integrable models without the addition of new particles into the spectrum.  As such, for a fixed spectrum only very special points correspond to integrable theories.  

Finally, let us now connect with the results of our companion paper \cite{PaperA}.  There we introduced the two spectra designated as follows:
\begin{itemize}
\item{Scenario I:} $S$ has a single pole corresponding to a particle of mass $m<m_{1}<2m$ and then a gap until $2m$.    
\item{Scenario II:} $S$ has a pole due to a cubic coupling (i.e. $m_{1}=m$) and then a gap until a heavier particle at $m_{2}$.  Between $m_{2}$ and $2m$ we place no restrictions on the spectrum. \end{itemize}
Scenario I is clearly the $N=1$ case considered above and which we also studied numerically in section \ref{DispSec} (see figures \ref{gvsm1} and \ref{SnumSCDD}).  Scenario II is slightly more subtle.  It turns out that when $m_{2}/m>\sqrt{3}$ it is equivalent to case $A$ of the $N=2$ example that we just considered in detail above (see figure \ref{optimizing_zero}a).  This seems counterintuitive at first sight since in the example considered above we explicitly allow for only the $m_{1}=m$ and $m_{2}$ poles below $2m$, while in Scenario II we impose no condition between $m_{2}$ and $2m$.   The equivalence is due to the fact that a CDD zero (when it must be added) will always cancel a pole.   To see this, consider adding an additional s-channel pole above $m_{2}$.  We see from \ref{optimizing_zero}a that this would mean that we have two consecutive s-channel poles so that we must insert a zero between them.\footnote{We are in case $A$ of figure \ref{optimizing_zero}a for $m_{2}>\sqrt{3}m$ since  $m_{1}/m=1< \sqrt{2} < \sqrt{3}< m_{2}/m$ and $m_{1}^{2}>4m^{2}-m_{2}^{2}$.  Note that $\th=0$ corresponds to the threshold at $s=4m^{2}$ so that in this case the pole closest to threshold is an s-channel pole.}  Optimizing the position of this zero would then cancel the new pole that we just added!  By the same argument poles corresponding to any number of particles heavier than $m_{2}$ would be canceled (so long as we do not allow for lighter particles which could produce t-channel poles above $m_{2}$).  
Thus we see that there is no need to impose a restriction above $m_{2}$ -- optimizing will always kill any poles corresponding to heavier particles.  In figure \ref{ANAvsCFT} we compare the numerical results obtained for these two scenarios in \cite{PaperA} with analytical results obtained here.  We see that the results obtained by these two very different means are in stunning agreement!

\section{The Ising Model with Magnetic Field}\label{IsingSec}

Figures \ref{TwoMasses3} and \ref{ThreeMasses2} are examples of bounds on couplings of a quantum field theory given some mass spectrum. An obvious question is whether there are interesting field theories saturating these bounds. Also, when the answer is \textit{no} what can we do to lower the bounds further until the answer is \textit{yes}? 
\begin{figure}[t!]
\begin{center}
\includegraphics[scale=0.6]{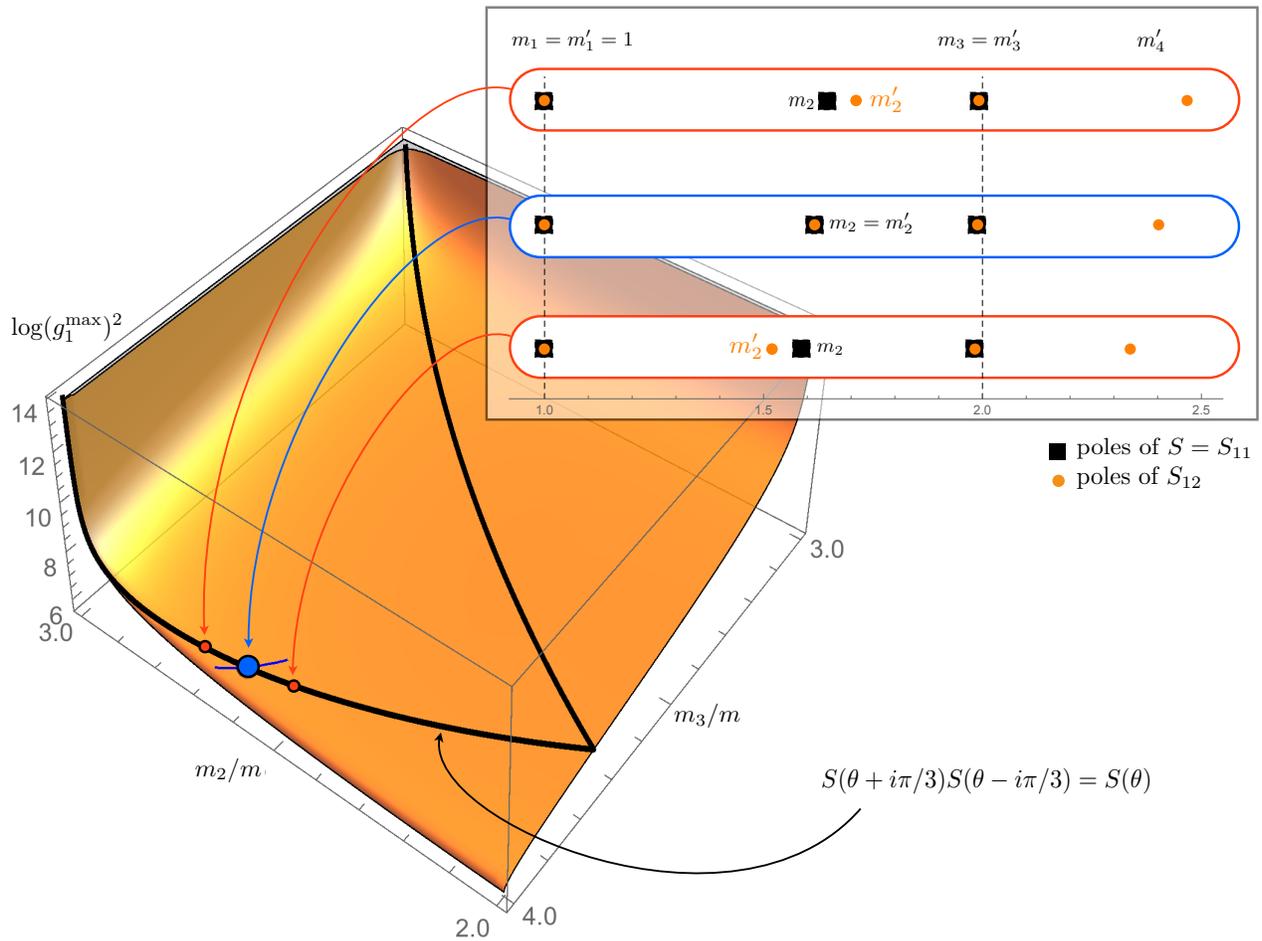} 
\end{center}
\caption{Blow up of region B from figure \ref{ThreeMasses2}.  The thick black line is where the cubic fusion property \eqref{cubic fusion} holds (i.e. assumption (1) in the discussion of section \ref{IsingSec}). In the upper right corner we plot the s-channel poles of $S_{12}$ versus those of $S$. We see that, following the thick black line, only at the blue dot does $S_{12}$ have poles at the same locations as $S$ indicating that assumption (3) from section \ref{IsingSec} also holds. } \label{IsingFigure}
\end{figure}

In some regions of these plots we already know the answer to these questions. Take for example the $(m_2/m)^2= 4$ section of figure \ref{TwoMasses3}. As $m_2\to 2 m$ this particle enters the two-particle continuum thus disappearing from the spectrum. We are thus left with a single exchanged particle $m_1$. This was precisely the case discussed in the simple numerics example and depicted in figure \ref{gvsm1}. For any $m_1>\sqrt{2} m$ we do know of a theory which saturates this bound: it is the Sine-Gordon integrable theory when we identify $m$ as the first breather and $m_1$ as the second breather. 

What about the more  general bounds in figures \ref{TwoMasses3} and \ref{ThreeMasses2}? All the optimal S-matrices  which maximize $g_1$ saturate unitarity and thus admit no particle production. Do they correspond to proper S-matrices of good integrable quantum field theories with their respective mass spectra? We will now argue that the answer to this question is \textit{no}. 

As an example we will focus on region $B$ in figure \ref{ThreeMasses2}. That is we will focus on the space of theories where there are three stable particles: the lightest particle itself with $m_1=m$ and two other heavier particles with 
\beq\label{ising region}
\sqrt{2} m_1 < m_2 < \sqrt{3} m_1 <m_3<2m_1 \,.
\eeq
In this region the S-matrix which maximizes $g_1$ is a simple product of three CDD factors, 
\beq
S(\theta)= \frac{\sinh(\theta)+i \sin(2\pi/3)}{\sinh(\theta)-i \sin(2\pi/3)}\times \frac{\sinh(\theta)+i \sin(\alpha_2)}{\sinh(\theta)-i \sin(\alpha_2)}  \times  \frac{\sinh(\theta)+i \sin(\alpha_3)}{\sinh(\theta)-i \sin(\alpha_3)} \,,\qquad m_j=2\cos(\alpha_j) \,.\la{SB}
\eeq
We will now argue that in the region \eqref{ising region} of parameter space our bound should \textit{not} be the strongest possible bound except at a single isolated point which we will identify with a well known and very interesting field theory.\footnote{The reader fond of section titles probably guessed which one.} We will do this by observing some simple pathologies with (\ref{SB}) which are resolved once $\alpha_2$ and $\alpha_3$ take some particular values which we identify below. 

To proceed we need to make three natural assumptions about a putative theory living in the boundary of our bounds for a fixed mass spectrum $\mathbb{M}$: 
\begin{enumerate}
\item[A1] The theory is integrable.\footnote{This is of course very natural since the S-matrices we found saturate unitarity and thus admit no particle production. Absence of particle production is of course a necessary condition for integrability. In most cases it is also a sufficient condition, see e.g. \cite{Iagolnitzer:1977sw}.}
\item[A2] The exchanged particle with mass $m_1=m$ is really the same as the external particle, i.e. it is not just another particle in the theory with the same mass as the external particle.
\item[A3] There are no other stable particles below the two particle threshold $2m_1$ other than those in $\mathbb{M}$.
\end{enumerate}  

\begin{figure}[t!]
\begin{center}
\includegraphics[scale=1.5]{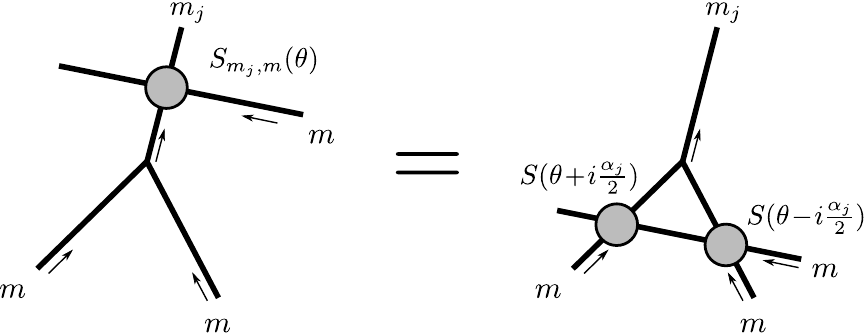} 
\end{center}
\caption{Suppose we take two (to be) constituents of a bound-state and throw them very slowly at each other so that they travel (almost) parallel to each other in space-time until they are close enough to feel each other and thus form the bound-state. Now suppose we want to scatter a fundamental particle with this bound-state as indicated on the left in this figure. This is how we would compute the left hand side of (\ref{BSS}). In an integrable theory we can shift at will the position of the wave packet of this fundamental particle. So we can shift it far into the past such that it scatters instead with the constituents of the bound-state well before they were bound together as represented on the right. This leads to the right hand side of (\ref{BSS}).  } 
\label{fusion}
\end{figure}

In an integrable theory we can construct bound-state S-matrix elements from the fundamental S-matrix by \textit{fusion}. If the stable particle shows up as a pole at $\theta=i \alpha_j$ in $S(\theta)$ then the S-matrix of this bound-state with the fundamental particle of mass $m$ can be built by scattering both its constituents \cite{Dorey:1996gd}, 
\beq
S_\text{j$^\text{th}$ bs, fund}(\theta)=S(\theta+i \alpha_j/2)S(\theta-i \alpha_j/2) \,. \la{BSS}
\eeq
This relation can be easily established starting with the $3\to 3$ ~S-matrix which is factorized as a product of three two-body S-matrices. We can then take two of the three particles in the initial state and form a bound-state. This will then describe a scattering of that bound-state with the remaining fundamental particle. (Because the theory is integrable, the individual momenta in the out state are the same as in the in-state so automatically we will be fusing into another bound-state in the future.) In this fusion process one of the three S-matrices (the one involving the particles being fused into a bound-state) simplifies (it yields a single pole of which we extract the residue) leaving us with two S-matrices which are nothing but the right hand side of (\ref{BSS}).  We can also justify (\ref{BSS}) in a more physical way as depicted in figure \ref{fusion}. 

With the fusion property \eqref{BSS} following from assumption A1 we will now show that powerful constraints on the spectrum follow from assumptions A2 and A3.

If a theory has a cubic coupling and $m_1=m$ shows up as a pole in the S-matrix then it can itself be thought of as a bound-state. That is, under the assumptions (1) and (2) above we conclude that we must have
\beq\label{cubic fusion}
S(\theta)=S(\theta+i \pi/3)S(\theta-i \pi/3) \,.
\eeq
This is an important self-consistency constraint. We can now plug the solution (\ref{SB}) in this relation. We observe that it is generically \textit{not} satisfied. However, there is a line $\alpha_3(\alpha_2)$ or equivalently $m_3(m_2)$ where it holds. This is the thick black line in figure \ref{IsingFigure}. Away from this black line we can already conclude that our bound is either not the optimal bound \textit{or} some of the assumptions A1 or A2 (or both) should not hold.

Sticking to the black line and continuing with assumption (3) we can do even better. We can now construct the S-matrix element~$S_{12}(\theta)=S(\theta+i \alpha_2/2)S(\theta-i \alpha_2/2)$ for the scattering~$m_1+m_2\to m_1+m_2$ involving the lightest and the next-to-lightest particles. We can then look at the poles of this S-matrix which will correspond to asymptotic particles of the theory. There is a point in the black line, marked with the blue dot in figure \ref{IsingFigure} where these poles correspond perfectly to the spectrum $\mathbb{M}=\{m_{1}(=m),m_{2},m_{3}\}$. Namely we find precisely three s-channel poles at $s=m_1^2,m_2^2,m_3^2<(2m_1)^2$ which are the very same locations in the fundamental S-matrix $S(\theta)$. However, as we move away from this blue point something bad happens. We see that the poles at $s=m_1^2$ and $s=m_3^2$ are as expected however the pole at $m_2^2$ shifts to a nearby position $m_2'^2$. This would indicate the presence of a new particle not in $\mathbb{M}$ with a mass close to that of $m_2$. This violates assumption A3. 

Ultimately, only the blue dot in figure \ref{IsingFigure} which is located at
\beq
m_2= 2\cos(\pi/5) m_1 \,, \qquad  m_3 =2\cos(\pi/3) m_1,
\eeq 
survives! We conclude that under the assumptions A1--A3 the maximal coupling in region B of figure \ref{ThreeMasses2} (which corresponds to masses satisfying \eqref{ising region}) should be lower than the one we found everywhere except perhaps at the blue point.\footnote{Note that we can not exclude having other integrable theories living in the black line provided we accept more stable particles below threshold showing up in other S-matrix elements. {We could also drop assumption A2 and conceive integrable theories where $m_1$ is not the same particle as the external one (despite having the same mass). If we keep assumption A3, the conclusion leading to the blue dot as a special isolated theory still holds.}}

What about this blue dot? Is there a special integrable theory with these masses and an S-matrix given by (\ref{SB})? \textit{Yes}, it is the Scaling Ising model field theory with magnetic field \cite{Zamolodchikov:1989fp}. This is a very interesting strongly coupled integrable theory with E8 symmetry which describes the massive flow away from the critical Ising model when perturbed by magnetic field (holding the temperature fixed at its critical value).\footnote{This is perhaps not that surprising. After all, many of the conditions we just imposed are simple recast of standard integrable bootstrap logic as used, for instance, in \cite{Zamolodchikov:1989fp}.} Thus the CDD solution provides a sharp (i.e. as strong as possible) upper bound on $g_1$ for this value of the masses. In what follows we shall refer to the blue dot in figure \ref{IsingFigure} as the \textit{magnetic} point. 

The thin blue line in figure \ref{IsingFigure} represent the variation of the masses of the stable particles $m_2$ and $m_{3}$ of the scaled Ising model as we move away from the magnetic point by shifting the Ising model temperature away from its critical value. The slope $\delta m_{2}/\delta m_{3}$ defining this line can be computed using so-called form factor perturbation theory as recalled in appendix \ref{FFSec}. As we change the temperature the corresponding field theory is no longer integrable (see \cite{Delfino:2003yr} for a review of the scaling Ising model with temperature and magnetic field turned on). Particle creation shows up to linear order in the thermal deformation but since this same particle production only shows up quadratically on the right hand side of (\ref{unitarityComplete}), its effect of the elastic component $S$ should be subleading. As such we expect that our bound for $g_{1}$ also captures the residue of the Scaled Ising model in the vicinity of the magnetic point. This is what we check in detail in appendix \ref{FFSec}.

 A conclusion of the discussion above is that   away from the magnetic point, the bound in figure \ref{IsingFigure} is not optimal. The obvious question is then how to improve it? 
One strategy would be to include other S-matrix elements into our analysis. In particular, it would be very interesting to consider the simplest absorptive components which are the inelastic $2\to2$ processes $m+m\to m+m_{2}$ and $m+m\to m_{2}+m_{2}$. Their existence, away from the integrable magnetic point, will forbid us to saturate unitarity for~$S(\theta)$ since they will show up in the right hand side of (\ref{unitarityComplete}). By taking them into account we expect therefore to be able to improve our bound.\footnote{Exactly at the Ising magnetic point, these inelastic amplitudes vanish due to a remarkable cancellation between poles associated with on-shell exchanged particles and 1-loop Coleman-Thun poles \cite{Dorey:Thesis}.  This mechanism was also noticed in the context of the sine-gordon model \cite{Goebel:1986na}.} As we add these components to our analysis, it would be formidable if a ridge-like feature passing the magnetic point represented by the blue dot would develop in figure \ref{IsingFigure}. By moving along this ridge we would hopefully be moving along the non-integrable thermal deformation thus accessing the full Scaling Ising model with temperature \textit{and} magnetic field. We are currently studying this problem and hope to report on progress in this direction in the near future. In the CFT bootstrap, adding further components to the analysis proved to be a very powerful idea \cite{Kos:2014bka}. Hopefully the same will be true here. It would also be very interesting to consider multi-particle scattering such as $2\to 3$ processes but these are kinematically more complicated and we did not dare explore them yet.

\section{Discussion}
\begin{figure}[t!]
\begin{center}
\includegraphics[scale=1.1]{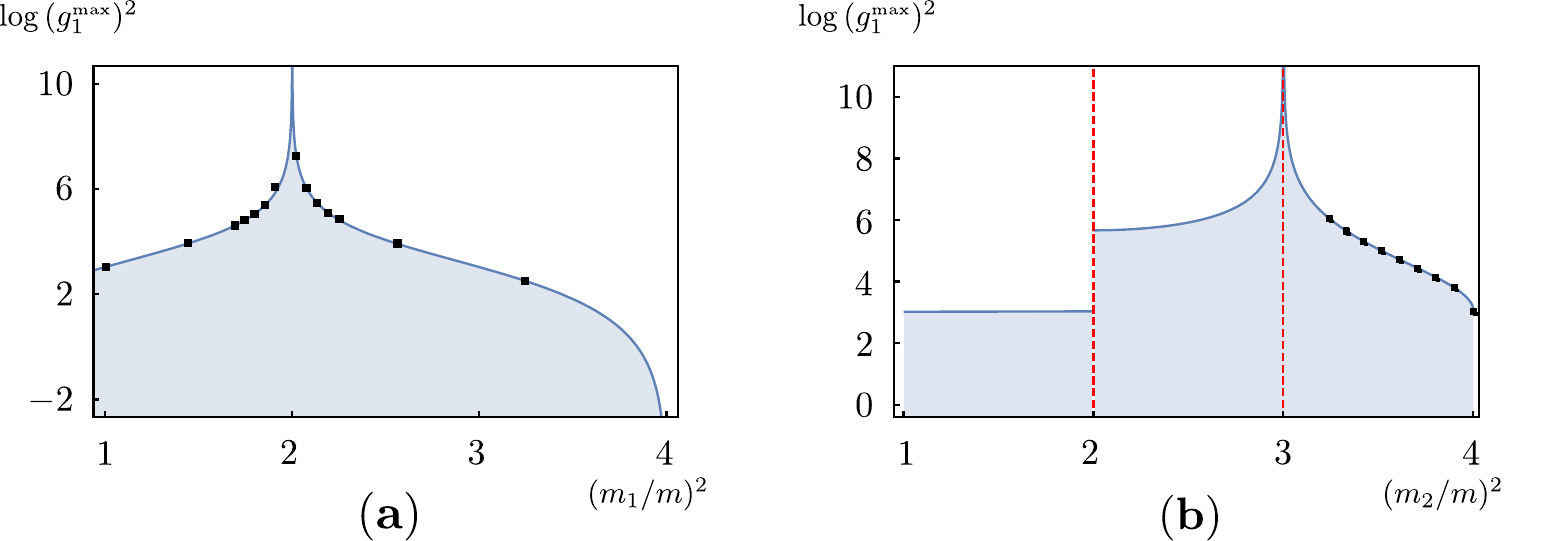} 
\end{center}
\caption{Maximal coupling $g_1^\text{max}$ for ({\bf a}) a single exchanged particle of mass $m_1$ and ({\bf b}) a particle of mass $m_1=m$ plus an heavier particle of mass $m_2$. The solid blue lines are the analytic results of the two dimensional S-matrix bootstrap. These are nothing but the top right and top left slices of the more general figure \ref{TwoMasses3}. The black squares are the outcome of the one dimensional conformal bootstrap numerics from \cite{PaperA}. These numerics are obtained using \texttt{SDPB} in ({\bf a}) and \texttt{JuliBootS} in ({\bf b})  \cite{Simmons-Duffin:2015qma,Paulos:2014vya,El-Showk:2016mxr}. In either case, within the precision of the numerics, the agreement with the analytic result is striking. It is worth emphasizing that the solid curves are very non-trivial functions. The right-most branch of ({\bf b}), for instance, corresponds to the analytic result~$(g_1^\text{max})^2 = { {12 ( {x} (6 \sqrt{4- {x}^2}-\sqrt{3}  {x} ( {x}^2-4))+3 \sqrt{3})}/({ {x}^4-4  {x}^2+3})}$ with~$ {x}=m_2/m$.} 
\label{ANAvsCFT}
\end{figure}
Armed with the insights of the remarkable recent progress in the \textit{conformal bootstrap} and with the well ironed technology of the \textit{integrable bootstrap}, we revisited here the \textit{S-matrix bootstrap} program. We found bounds on the   maximal couplings in massive two dimensional quantum field theories with a given mass spectrum. We obtained these bounds numerically (see section \ref{DispSec}) and analytically (see section \ref{CDDsec}) with perfect agreement between the two methods, see e.g. figure \ref{gvsm1}. These bounds also made contact with well known integrable theories. We found, for example, that there is no unitary relativistic quantum field theory in two dimensions whose S-matrix element for identical particles has a single bound-state pole at $s=m_{1}^2>2m^2$ and a bigger residue than that of the Sine-Gordon breather S-matrix. 

In the companion paper~\cite{PaperA} we attacked this problem from yet a different perspective. There, we considered a Gedanken experiment where we put massive ($D$-dimensional) quantum field theories into a (Anti-de Sitter fixed background) box. We can then study their landscape by analyzing  the conformal theories they induce at the ($D-1$ dimensional) boundary of this space-time. This allows us to make use of well-developed numerical methods of the conformal bootstrap for putting bounds on conformal theory data which then translate into bounds on the flat space QFT data.
An important difference with respect to previous works on conformal bootstrap is that this setup  requires all conformal dimensions involved in the bootstrap to be very large. This is how we make sure the AdS box is large and the physics therein is equivalent to that in flat space. This poses significant technical challenges as discussed in detail in \cite{PaperA}. This method of extracting QFT bounds is very onerous and requires several hours of computer time whereas the numerical method described in this paper takes a few seconds. Beautifully, in the end, the two calculations match as illustrated in figure \ref{ANAvsCFT}.\footnote{There is a slight difference in notation w.r.t. to that paper. Here we use $m$ for the external particle and $m_1,m_2,\dots$ for the exchanged particles. In \cite{PaperA} we use $m_1$ for the external particle. For the exchanged particles we then use  $m_2$ for the case corresponding to the left plot in figure \ref{ANAvsCFT} while we denote them as $m_1$ and $m_b$ in the case corresponding to the right plot in figure \ref{ANAvsCFT} where there is a cubic self-coupling.}

We find the agreement between the conformal bootstrap and the S-matrix bootstrap to be conceptually very interesting. (At least in the case at hand corresponding to $D=2$) we observe that the $D-1$ dimensional conformal bootstrap \textit{knows} about the $D$ dimensional massive S-matrix bootstrap. From an AdS/CFT-like intuition this is perhaps to be expected since we can always put whatever we want into boxes. On the other hand, we still find it comforting albeit counterintuitive that we can learn about massive quantum field theories from conformal theories in one lower dimension. 


There are two natural follow up directions to this work and \cite{PaperA}. One is to explore further the two dimensional world by including into the analysis S-matrix elements involving heavier particles. When these other components do not vanish, unitarity is not saturated and therefore we expect in this way to make contact with interesting non-integrable theories. One may learn, for example, about the full scaling Ising model with magnetic and temperature deformations as discussed at the end of section \ref{IsingSec}.   The second promising direction would be to stick with the simplest S-matrix element involving identical lightest particles but move to higher dimensions. In both cases we no longer expect the luxury of analytic results as obtained here. The hope, however, is that proper generalizations of the numerical methods -- both the S-matrix and the conformal bootstrap one -- will survive. 

From the conformal bootstrap point of view, either direction is straightforward although technically challenging. The technology for dealing with multiple correlators exists \cite{Kos:2014bka} and going to higher dimensions also does not pose any conceptual issues. In either case we can however expect the numerical computations to become even more demanding than for two-dimensional QFTs.
From the perspective of the S-matrix bootstrap it seems simple to include amplitudes involving heavier particles. We are also optimistic about a similar analysis as the one of this paper but for higher-dimensional QFTs. We hope to report on progress in these directions in the near future. 

In any case,  it seems very fruitful to pursue the conformal and S-matrix bootstrap hand-in-hand.  
Both for the multiple correlator story as well as for higher dimensions, having a conformal bootstrap bound, even if it is numerically hard to get, would serve as an invaluable hint. Such lampposts are extremely valuable and may provide key insights to the S-matrix bootstrap which were missing in the 60's.

\emph{Note added:} When we were about to submit this paper we found a surprisingly unknown\footnote{It seems like we are the first ones to cite it!} 44 year old paper by Michael Creutz \cite{Creutz:1973rw} which further refers to a 55 year old book chapter by Symanzik \cite{nobodycanfindSymanzik}. In this beautiful two page paper many of the results of this paper are derived in a rather elegant way. We rediscovered here various of the arguments present there. The relation to the conformal bootstrap and the connection to various known integrable models pointed out in our work seems novel and so does the numerical approach -- which we believe can be extended to higher dimensions.

 \section*{Acknowledgements}

We thank Benjamin Basso, Patrick Dorey, Davide Gaiotto, Martin Kruczenski, Rafael Porto, Slava Rychkov, Amit Sever and Alexander Zamolodchikov for numerous enlightening discussions and suggestions. 
Research at the Perimeter Institute is supported in part by the Government of Canada through NSERC and by the Province of Ontario through MRI. 
MFP was supported by a Marie Curie Intra-European Fellowship of the European Community's 7th Framework Programme under contract number PIEF-GA-2013-623606.
Centro de F\'{i}sica do Porto is partially funded by FCT. 
The research leading to these results has received funding from the People Programme (Marie Curie Actions) of the European Union's Seventh Framework Programme FP7/2007-2013/ under REA Grant Agreement No 
 317089 (GATIS) and the grant CERN/FIS-NUC/0045/2015.
 JP is supported by the National Centre of Competence in Research SwissMAP funded by the
Swiss National Science Foundation.

\appendix

\section{Numerics}\label{numerics}
In this appendix we give more details on the numerics described in section \ref{DispSec}.  We consider a grid $\{x_0,x_1\dots x_{M}\}$ and measure everything in units of $m$ so that $x_{0}=4$.  Denote by $\rho_{n}$ the value $\rho(x_{n})$ and approximate $\rho(x)$ by a linear spline connecting the points $(x_{n},\rho_{n})$ as shown in figure \ref{linear_spline}.  We can then perform the integrals in \eqref{dispersion} analytically giving \eqref{discrete dispersion} with
\beqa
K_a(s)=\!\!\!\!\!\!\!\!&&\frac{\left(x_{a-1}-s\right) \log \left(x_{a-1}-s\right)}{x_{a-1}-x_a}+\frac{\left(x_{a+1}-s\right) \log
  \left(x_{a+1}-s\right)}{x_a-x_{a+1}}-\frac{\left(x_{a-1}-x_{a+1}\right) \left(x_a-s\right) \log
  \left(x_a-s\right)}{\left(x_{a-1}-x_a\right) \left(x_a-x_{a+1}\right)}\nn \\
  &&+\,(s\to 4-s) \nn\,,
\eeqa
with $a=1,2,\dots,M-1$ while for the last point of the grid
\beqa
K_M(s)=\!\!\!\!\!\!\!\!&&\frac{\left(x_{M-1}-s\right) \log \left(x_{M-1}-s\right)}{x_{M-1}-x_M}-\frac{x_M \log
  \left(x_M\right)}{s}+\frac{\left(x_{M-1}-x_M-s\right) \left(x_M-s\right) \log \left(x_M-s\right)}{s
  \left(x_{M-1}-x_M\right)}-1 \nn \\
  &&+\,(s\to 4-s) \nn\,.
\eeqa 
Note that for  $x>x_{M}$ we assume a tail of the form $\rho(x)\sim \rho_{M} \, x_{M}/x$ which leads to the above result for $K_{M}$.

We can now evaluate the approximate S-matrix \eqref{discrete dispersion} at a bunch of points with $s \ge 4$.  It is convenient to evaluate on the gridpoints $x_{a}$ themselves (although not necessary of course) so that we can make use of the identity
\beq
\text{Im} \, S(x_{a}+i0) = \pi \rho_{a}.
\eeq
This gives a set of $M$ constraints\footnote{Note that $\text{Re}[K_{n}(x_{a})]$ can be computed simply by replacing $\log(\dots)\to \log(\text{abs}(\dots))$ in the expressions for $K_{n}(s)$.} 
\beq
\[S_\infty-\sum_{j} \mathcal{J}_j \(\frac{g_{j}^{2}}{x_{a}-m_{j}^{2}}+\frac{g_{j}^{2}}{4m^2-x_{a}-m_{j}^{2}}\)+\sum_{n=1}^{M}\text{Re}\[K_{n}(x_{a})\] \rho_{n}\]^{\!\!2}+\(\pi \rho_{a}\)^{2}\le 1 
\label{discrete unitarity}
\eeq
for $a=1,...,M$.  The goal is, for a given set of masses $m_{j}$, to find the point in the space $\{S_{\infty},g_{1},g_{2},...,g_{N},\rho_{1},...,\rho_{M}\}$ such that $g_{1}$ is as big as possible and the constraints \eqref{discrete unitarity} are satisfied.
\begin{figure}[t!]
\begin{center}
\includegraphics[scale=1.0]{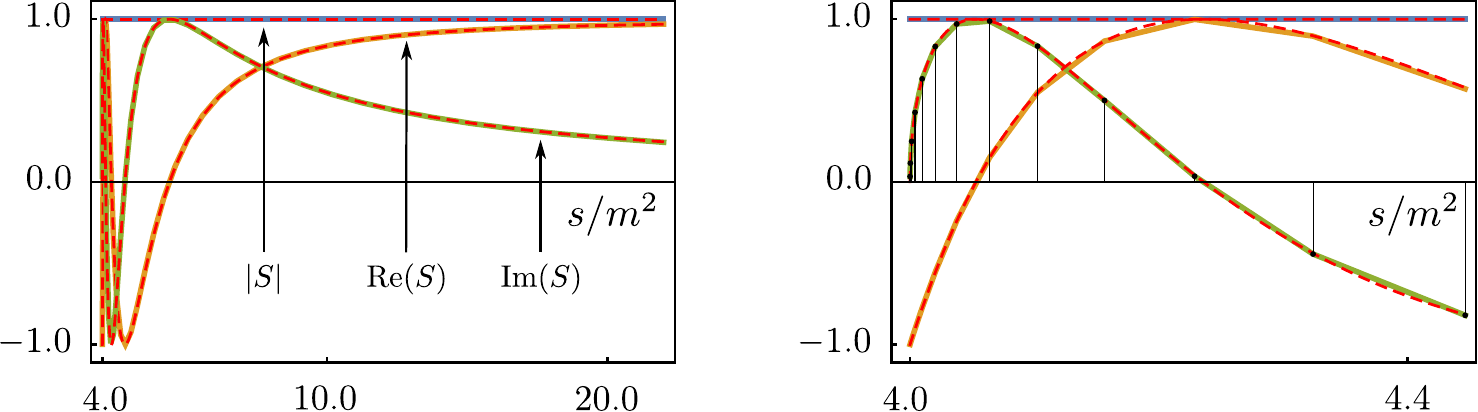} 
\end{center}
\caption{Result of numerics for the spectrum $\mathbb{M}=\{1,1.6,1.8\}$ compared with the expectation~\eqref{SB} and its near-threshold close-up (on the right). The green, orange and blue curves are $\text{Im}(S_{\text{num}})$, $\text{Re}(S_{\text{num}})$, $|S_{\text{num}}|$ where $S_{\text{num}}$ is \eqref{discrete dispersion} evaluated on the result of the numerics given in (\ref{outcome}). The black dots indicate the points $(x_{n},\rho_{n})$; note that we use a grid which clusters points near threshold.  The dashed red curves are the corresponding parts of the exact solution \eqref{SB}.  } 
\label{SnumSCDD_appendix_example}
\end{figure}
This amounts to a standard problem in quadratic optimization and the \texttt{Mathematica} program \texttt{FindMaximum} is conveniently designed to carry out such a task.  The attached notebook contains our implementation of this problem in Mathematica.  There we implement a function \texttt{MaxCoupling[$\mathbb{M}\_$]} which takes a spectrum $\mathbb{M}$ as input and returns the maximum value of $g_{1}$ along with the corresponding values of the variables $g_{j>1}$, $\rho_{n}$ and $S_{\infty}$.  To illustrate with a typical example, the output of $\texttt{MaxCoupling}$ for $\mathbb{M}=\{1, 16/10,18/10\}$ (in units of~$m$) is 
%
\beq
\includegraphics[scale=0.48]{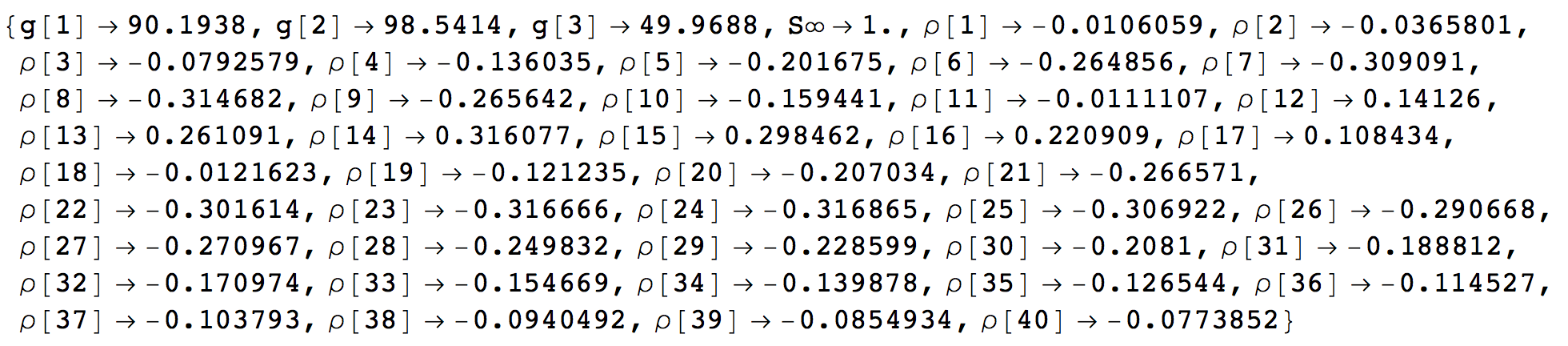} \label{outcome}
\eeq
Note that this is within the parameter range \eqref{ising region} which is the region plotted in figure \ref{IsingFigure} and also region B in figure \ref{ThreeMasses2}.  Thus we expect the S-matrix to be given by \eqref{SB} with the values of $\a_{j}$ chosen according to $\mathbb{M}$.  We can see in figure \ref{SnumSCDD_appendix_example} that our numerics agree perfectly with expectation.

\section{Hyperbolic Rapidity} \label{thetaAp}
In two dimensions, hyperbolic rapidities are a very useful parametrization of energy and momenta of relativistic particles. For two particles, for instance, we would write
\beq
p_1^\mu=(E_1,\vec{p}_1)=(m_1 \cosh(\theta_1),m_1 \sinh(\theta_1)) \,, \qquad p_2^\mu=(E_2,\vec{p}_2)=(m_2 \cosh(\theta_2),m_2 \sinh(\theta_2)) \,.\nn
\eeq
In this parametrization consider the elastic scattering of these two particles. In the final state, conservation of energy and momentum imply that the final individual momenta are the same as the initial one, that is $p_3=p_1$ and $p_4=p_2$ so that there is no momentum exchange $u=(p_3-p_1)^2=0$. As for the other Mandelstam invariants we have
\beq
s=(p_1+p_2)^2=m_1^2+m_2^2+2m_1m_2 \cosh(\theta) \,, \qquad t=(p_2-p_3)^2=m_1^2+m_2^2-2m_1m_2 \cosh(\theta) \,, \nn
\eeq
where $\theta=\theta_1-\theta_2$ is the difference of hyperbolic rapidities. 

Note that these relativistic invariants are invariant under shifts of both rapidities. Indeed, boosts act as shifts of $\theta_1$ and $\theta_2$ such that $\theta$ is invariant. 

Note also that $\theta \leftrightarrow i\pi-\theta$ is a crossing transformation which exchanges $s$ and $t$. This is also nicely seen directly in terms of the two vectors above. For instance, if we keep $\theta_2$ fixed and send $\theta_1 \to i\pi-\theta_1$ we get that $p_1^\mu\to (-E_1,\vec{p}_1)$ as expected for a crossing transformation. This sends $p_1$ to the future and $p_3$ to the past.

The hyperbolic parametrization is also convenient when dealing with bound-states. Suppose for instance we form a bound-state out of two constituent particles with rapidities $\theta \pm i \eta$ and mass $m$. Then the total two-momenta of the bound-state would be 
\beqa
p_\text{bound-state}&=&(m \cosh(\theta+i\eta) + m \cosh(\theta-i\eta) , m \sinh(\theta+i\eta) + m \sinh(\theta-i\eta)  ) \nn \\
&=&(m_\text{bound-state} \cosh(\theta),m_\text{bound-state} \sinh(\theta)) \,,
\eeqa
where the bound-state mass
\beq
m_\text{bound-state}= 2m \cos(\eta)  \,.
\eeq
A necessary (but not sufficient) condition for such bound-states to form is the existence of a pole at $\theta=2i\eta$ in the S-matrix element $S(\theta)$ describing the elastic process $m+m\to m+m$. 

Some theories have a cubic coupling and the particle of mass $m$ can also be though of a bound-state of two particles of mass $m$. In these cases $\eta=\pi/3$. The Ising field theory with magnetic field discussed in the main text is one such example. 

\section{Form Factor Expansion} \la{FFSec}

The so called Scaled Ising Field Theory is a remarkable field theory, see \cite{Delfino:2003yr} for a beautiful review. This theory describes the flow from the critical Ising model as we turn on magnetic field and temperature (measured as a deviation from its Curie value). When we turn on temperature only (without magnetic field) or magnetic field alone (without temperature) we end up with integrable theories. The first is that of free fermions while the second is Zamolodchikovs E8 theory \cite{Zamolodchikov:1989fp}. We rediscovered this second special point in section \ref{IsingSec} as the integrable theory with three stable particles of masses in the range (\ref{ising region}) and whose cublic coupling to the lightest particle is maximal. 

Away from these two Integrable points, the Scaled Ising Field Theory can be studied numerically, either from the lattice or using the so-called Truncated Conformal Space Approach \cite{Fonseca:2001dc, Fonseca:2006au, Zamolodchikov:2013ama}. We can also use Integrable Form Factor perturbation \cite{Delfino:1996xp, Delfino:2005bh} theory to study small deformations away from the integrable points. Let us discuss briefly how our general bounds in figure \ref{IsingFigure} compare with this second analysis. 

As we deform away from the E8 theory by turning on the temperature slightly the masses of the stable particles move. More precisely, we chose to measure everything in terms of the mass of $m_1=1$ which is thus kept fixed but $m_2$ and $m_3$ will move. This is a slightly different point of view compared to what is typically taken in the literature -- see e.g. \cite{Delfino:2005bh} -- where masses are measured in unit of magnetic field. In this convention all masses move as we deform away from the integrable point. The results (in this second notation) are given in equations (11) and (64) of \cite{Delfino:2005bh}. Converting to our conventions we get therefore
\beq
\left.\frac{\delta m_2}{\delta m_3} \right|_\text{here}= \left.\frac{\delta (m_2/m_1)}{\delta (m_3/m_1)} \right|_\text{there} = \left.\frac{\delta m_2 - m_2/m_1\, \delta m_1 }{\delta m_3 - m_3/m_1\, \delta m_1} \right|_\text{there}  \simeq 1.57322 \,.
\eeq
In the small thin blue line in figure \ref{IsingFigure} we marked this slope. We can now compute the slope of our bound for $g_1^\text{max}$ along this blue line. We find
\beq
\log(g_1)=6.585891698-8.683281573 \, \delta m_2+ O\left((\delta m_2)^2\right) \,. \la{expect}
\eeq
This value must coincide with the variation of the coupling of the Scaled Ising model as we move away from this point or else we will violate our bound as we slightly increase or decrease the temperature. This is what we verified. It is a somehow involved check since extracting this residue from the form factor expansion is considerably harder than correcting the masses. Fortunately, 
attached to \cite{Delfino:2005bh} is a long notebook with the four-particle form factor for the energy density operator. Using it we can construct the correction $\delta S(\theta)$ to the S-matrix. From it, we can read off the correction $\delta g_1$ to the cubic coupling to the lightest particle. In this way we obtain exactly the slope (\ref{expect})  (within the eleven digits of numerical precision of the notebook in \cite{Delfino:2005bh}). 

We further checked that the S-matrix as extracted from \cite{Delfino:2005bh} is in fact still of CDD form to first order in the deformation. (In checking this it is important to shift $\theta$  appropriately as to preserve the standard relation $s(\theta)$ and thus maintain crossing in its usual form.) This is expected in fact since, as mentioned in the main text, particle production is produced to leading order in the deformation but this particle production shows up quadratically in the unitarity constraint thus only inducing absorption in the elastic S-matrix element to quadratic order.  

\section{Most General Optimal CDD solution}\label{CDDgeneral}

A given mass spectrum $\{m_1/m,\dots,m_N/m\}$ leads to $2N$ poles between $\theta=0$ and $\theta=i\pi$. They come in pairs (for the s-channel and the t-channel contribution) related by $\theta \leftrightarrow i\pi-\theta$ and thus it is enough to focus on the segment $[0,i\pi/2]$. We order the poles in this segment and denote them as $\theta_j=i\gamma_j$ with $0<\gamma_1<\gamma_2 < \dots <\gamma_N<\pi/2$.  (Needless to say, this ordering is \textit{not} the same as the order $m_1<m_2<\dots<m_N$.) To each pole~$\gamma_j$ we associate a sign $\text{sgn}(j)=+1$ if this is an s-channel pole or $\text{sgn}(j)=-1$ for a t-channel pole. In this way, the set $\{(\gamma_1,\text{sgn}(1)),\dots,(\gamma_N,\text{sgn}(N))\}$ encodes all the information about the mass spectrum. 
In terms of this useful notation, the optimal solution is simply
\beq
S_{g_1^\text{max}}(\theta)=\text{sgn(1)} (-1)^{N-1} \times \prod_{j=1}^{J-1} [\gamma_j]^{\frac{1-\text{sgn}(j)\text{sgn}(j+1)}{2}} \times \[\gamma_J\]  \times \prod_{j=J+1}^{N} [\gamma_j]^{\frac{1-\text{sgn}(j-1)\text{sgn}(j)}{2}} \la{optimal}
\eeq
where $\gamma_J$ is the pole associated to the lightest exchanged particle, that is $m_1^2=2m^2 (1+\text{sgn}(J) \cos(\gamma_J))$ or $m_1^2(4m^2-m_1^2)=4m^2 \sin^2(\gamma_J)$.  In words, the optimal solution (\ref{optimal}) carefully removes CDD-poles whenever the alternating pattern between s- and t-channel poles is not observed. This immediately guarantees that all signs come out right. The optimal residue $g_1^\text{max}$ can now be straightforwardly read from (\ref{near pole of S}) and (\ref{optimal}) as 
\beq
\(g_1^\text{max}\)^2= 16 \sin^2\gamma_J \times \(\Gamma_1^\text{max}\)^2
\eeq
with
\beq
\(\Gamma_1^\text{max}\)^2 = \sigma_1 (-1)^{N-1} \, 2  \tan(\gamma_J) \, \prod_{j=1}^{J-1} \(\frac{\sin\gamma_J+\sin\gamma_j}{\sin\gamma_J-\sin\gamma_j}\)^{\!\!\frac{1-\sigma_j \sigma_{j+1}}{2}} \!\!\! \prod_{j=J+1}^{N} \!\!\(\frac{\sin\gamma_J+\sin\gamma_j}{\sin\gamma_J-\sin\gamma_j}\)^{\!\!\frac{1-\sigma_j \sigma_{j-1}}{2}}  \la{optimalGamma}
\eeq
where we are using the shorthand $\s_j\equiv \text{sign}(j)$.


\end{document}